\documentclass[acmtog,screen]{acmart}

\usepackage{booktabs} 
\usepackage[utf8]{inputenc}
\usepackage[vlined,ruled]{algorithm2e}
\usepackage{subcaption}
\usepackage{amsmath}
\usepackage{lipsum} 
\usepackage{multirow}
\usepackage{mathtools}
\usepackage{wrapfig}
\usepackage{enumitem}
\setlist{nosep}

\setcitestyle{square}

\SetAlFnt{\small}
\SetAlCapFnt{\small}
\SetAlCapNameFnt{\small}
\SetAlCapHSkip{0pt}
\IncMargin{-\parindent}

\acmJournal{TOG}
\acmYear{2021}

\setcopyright{usgovmixed}

\acmDOI{0000001.0000001_2}

\begin{document}
\title{Neural UpFlow: A Scene Flow Learning Approach to Increase the Apparent Resolution of Particle-Based Liquids}

\author{Bruno Roy}
\affiliation{
  \institution{Université de Montréal}
  \country{Canada}
}
\affiliation{
  \institution{Autodesk}
  \country{Canada}
}
\email{bruno.o.roy@umontreal.ca}


\author{Pierre Poulin}
\affiliation{
  \institution{Université de Montréal}
  \country{Canada}
}
\email{poulin@iro.umontreal.ca}

\author{Eric Paquette}
\affiliation{
 \institution{École de technologie supérieure}
 \country{Canada}
}
\email{eric.paquette@etsmtl.ca}

\renewcommand\shortauthors{Roy, B. et al.}

\begin{abstract}
We present a novel up-resing technique for generating high-resolution liquids based on scene flow estimation using deep neural networks. Our approach infers and synthesizes small- and large-scale details solely from a low-resolution particle-based liquid simulation. The proposed network leverages neighborhood contributions to encode inherent liquid properties throughout convolutions. We also propose a particle-based approach to interpolate between liquids generated from varying simulation discretizations using a state-of-the-art bidirectional optical flow solver method for fluids in addition with a novel key-event topological alignment constraint. In conjunction with the neighborhood contributions, our loss formulation allows the inference model throughout epochs to reward important differences in regard to significant gaps in simulation discretizations. Even when applied in an untested simulation setup, our approach is able to generate plausible high-resolution details. Using this interpolation approach and the predicted displacements, our approach combines the input liquid properties with the predicted motion to infer semi-Lagrangian advection. We furthermore showcase how the proposed interpolation approach can facilitate generating large simulation datasets with a subset of initial condition parameters. 
\end{abstract}



\begin{teaserfigure}
    \centering
    \begin{subfigure}[h]{0.33\textwidth}
        \centering
        \includegraphics[width=\linewidth]{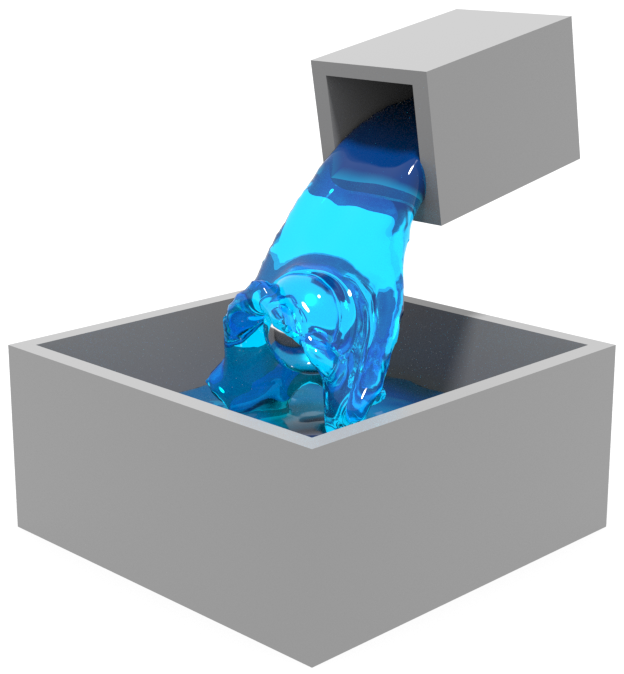}
    \end{subfigure}
    \begin{subfigure}[h]{0.33\textwidth}
        \centering
        \includegraphics[width=\linewidth]{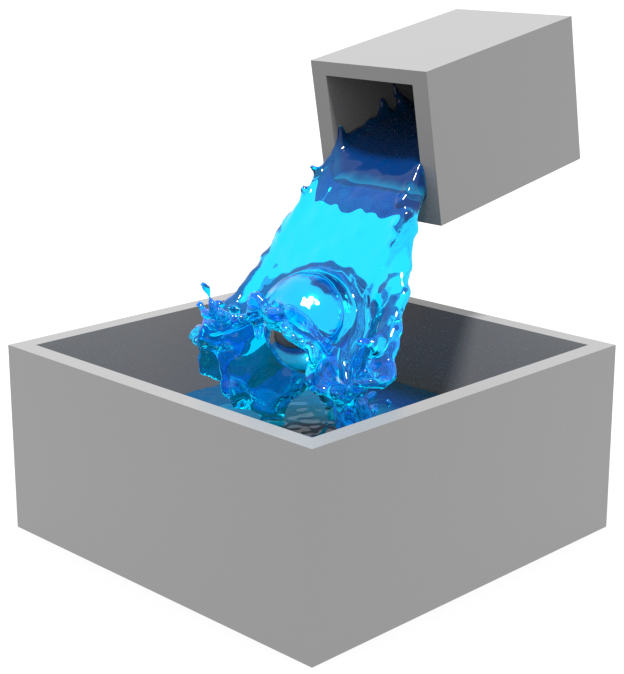}
    \end{subfigure}
    \begin{subfigure}[h]{0.33\textwidth}
        \centering
        \includegraphics[width=\linewidth]{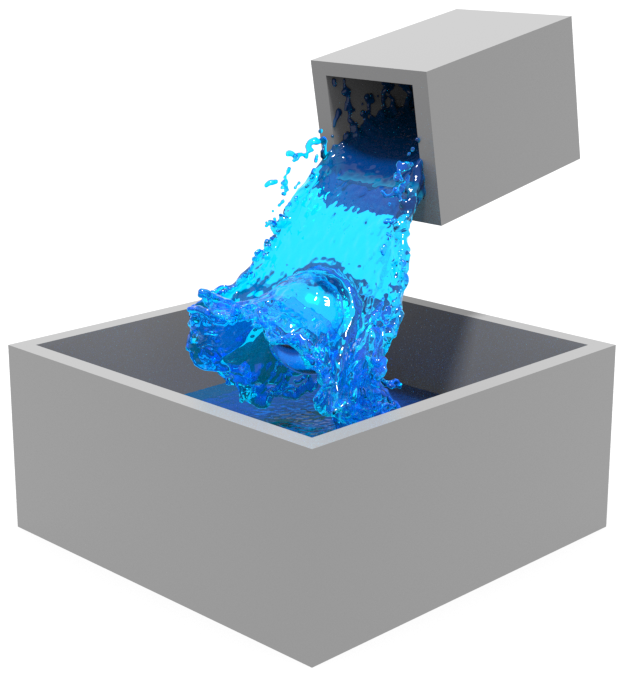}
    \end{subfigure}
    \caption{Our learning deformation field reproduced~(middle) most of the small- and large-scale details from the high-resolution ground truth~(right) using solely the low-resolution input example~(left). Particles are added and displaced by our network to generate \textit{plausible} details of up-resed animations at a fraction of a high-resolution simulation cost.}
    \label{fig:teaser}
\end{teaserfigure}

\maketitle

\section{Introduction}
Machine learning approaches have been rising in popularity in the last few years due to their accessibility and versatility. These learning approaches have been adapted for many applications in media and entertainment areas, and more specifically in graphics, such as animation~\cite{aberman2020skeleton}, motion capture~\cite{tung2017self}, style transfer~\cite{wang2016unsupervised}, 2D-to-3D sketching techniques~\cite{delanoy20183d}, physically-based rendering~\cite{chaitanya2017interactive}, texture synthesis~\cite{gatys2015texture}, fluid simulations~\cite{kim2019deep}, and so on.

In recent years, the use of machine learning on fluids has considerably reduced the emphasis on computationally expensive numerical methods while taking advantage of the existing knowledge in the field. So far, machine learning techniques have been used to improve numerical solvers~\cite{ladicky2015data, yang2016data, tompson2017accelerating}, to compute fluid features~\cite{ummenhofer2019lagrangian, wiewel2019latent, kim2019deep, um2018liquid}, and to infer visual details~\cite{prantl2018generating, chu2017data}. However, very few of these methods have directly addressed the notion of improving the apparent resolution of \textit{particle}-based fluids, particularly in the context of learning on unstructured data. Lately though, researchers in the computer vision community~\cite{qi2017pointnet, qi2017pointnet++} have studied irregular data, such as point clouds, and introduced ways to directly process them as inputs for 3D classification and segmentation tasks. Moreover, their work has been recently extended~\cite{liu2019flownet3d, wang2020flownet3d++} to learn on temporal motions of point clouds such as defined using scene flow estimation methods. Considering structural similarities between point clouds and particle systems, we demonstrate in this paper that using scene flows expressed as Lagrangian motions on particles is promising to infer fine and controllable details on fluids.

The work most similar in spirit to ours is the one by Prantl et al.~\shortcite{prantl2018generating}. They propose to leverage neural networks to learn precomputed deformations of liquid surfaces. While we partially share their goal in encoding space-time deformations for liquids, there are several noticeable differences. First, we target particles as opposed to signed distance functions (SDF) to provide greater flexibility on the method used to apply deformations. We show that applying deformations directly on particles before generating a surface facilitates bringing out fine details, often depending on the discretization. Second, using a neighborhood embedding layer, our network does not require multiple learning phases to encode realistic high-resolution liquid features. In addition, we use a deformation weight obtained from the work of Thuerey~\shortcite{thuerey2017interpolations} to guide (i.e., through our deformation-aware loss function) and speed up convergence.

In this paper, we propose an adapted deep neural network architecture that combines a recent scene flow machine learning method with particle neighborhood interactions expressed as fluid simulation properties. Our network exploits particle neighborhoods inspired by hybrid liquid simulation methods to encode hierarchical features of the particle-based simulations during the convolution operations. We show that we are able to preserve important liquid features at every level of convolution. Our loss function also includes a coefficient to weigh important local features while reducing convergence time. Finally, we introduce an inter-resolution interpolation framework that allows us to displace the particles while taking the actual low-resolution motion into account. From this point on, we refer to \textsl{inter-resolution} as a correspondence between a low-resolution simulation and a high-resolution simulation. This same interpolation framework is also used to artificially grow our existing multi-resolution simulation dataset. With this comprehensive learning pipeline, our approach is able of accurately reproducing high-resolution details using solely a very coarse particle-based liquid. To summarize, the main contributions of our work are:
\begin{itemize}[leftmargin=0.4cm]
\item{an adapted deep neural network architecture using particle
neighborhoods and a deformation-aware loss function reducing
the inference noise encoded during convolutional downsampling
operations,}
\item{an interpolation approach that transposes a scene flow into
Lagrangian motion using the input simulation properties, and}
\item{a data augmentation framework that artificially grows an
existing particle-based simulation dataset.}
\end{itemize}

\section{Related Work}

\paragraph{\textbf{Multi-Scale and Procedural Methods}}
In the last decade or so, several methods have been proposed to enhance the apparent resolution of liquid simulations. While procedural methods were more practical for decoupling simulations from discretization, multi-scale methods were particularly efficient to process independently surface details and separate them from coarse volumes of liquid. Using a resampling strategy to segregate multiple layers of the simulation data (e.g., particles) has been revisited multiple times. Adaptive models were proposed to dynamically adjust the size (and contributions on forces) of particles using split-and-merge operations within the simulation loop~\cite{adams2007adaptively, winchenbach2017infinite}. While these methods share the same goal in spirit, Winchenbach et al.~\shortcite{winchenbach2017infinite} leverage the fundamental basis of Smoothed Particle Hydrodynamics (SPH) methods. Their proposed scheme enables defining with precision particle masses to improve stability while preserving small-scale details. Similarly, Solenthaler and Gross~\shortcite{solenthaler2011two} introduced an adaptive method to couple a dual particle-layer scheme using two distinct particle resolutions. The method of Winchenbach et al.~\shortcite{winchenbach2017infinite} was also recently extended to improve spatial adaptivity by introducing a continuous objective function as a refinement scheme for SPH~\cite{winchenbach2021optimized}.

A method focusing on preserving thin sheets was introduced by~Ando et al.~\shortcite{ando2012preserving} allowing them to adapt the resolution of particles in \textsl{Fluid Implicit Particle (FLIP)} simulations. Later, narrow-band methods were introduced to get the best of both Eulerian and Lagrangian schemes by expressing the liquid surface with particles and the coarser volume with a grid~\cite{ferstl2016narrow}. The method was first introduced to FLIP given its semi-Lagrangian nature. The method was then extended by Sato et al.~\shortcite{sato2018extended} to process arbitrary locations (as opposed to exclusively the liquid surface), identifying where particles are needed to refine the appearance of the surface. The same idea was revisited and adapted to SPH methods~\cite{raveendran2011hybrid, chentanez2015coupling, roy2018hybrid}. Other adaptive methods were also proposed to exploit spatially adaptive structures to capture different simulation scales~\cite{ando2013highly, aanjaneya2017power}.

Meanwhile, procedural methods have also been introduced as a refinement scheme by generating surface points through wave simulations. As noted by Kim et al.~\shortcite{kim2013closest}, high-frequency details are not necessarily coupled to the coarse simulation and can be independently generated directly onto the animated mesh. In comparison to the work of Kim et al.~\shortcite{kim2013closest}, which was applied in an Eulerian setup (i.e., solely based on the corresponding SDF and the velocity field), Mercier et al.~\shortcite{mercier2015surface} proposed a Lagrangian up-res method to increase the apparent resolution of FLIP with a secondary surface wave simulation. Recently, another procedural method was introduced focusing solely on splashing liquid details~\cite{roy2020particle}. As we focus on neural networks (NN) to enrich the liquid surface, we will not compare our work with procedural methods. Although we share the similar goal of improving the apparent resolution of the liquid surface at a fraction of the cost and computational time, our approach offers the capabilities to synthesize plausible details because it is trained on real fluid simulations.

\paragraph{\textbf{Machine Learning for Fluids}}
Leveraging machine learning methods for fluids was first introduced in graphics by Ladick\`y et al.~\shortcite{ladicky2015data}. They demonstrated the inference of SPH forces to approximate Lagrangian positions and velocities using regression forests. Along the same line of ideas, a similar method was lately proposed to apply neural networks to Eulerian methods~\cite{yang2016data}. More recently, NNs were combined with numerical solvers to model diffuse particle statistics for the generation of splashes~\cite{um2018liquid}. Convolutional neural networks (CNN) were also widely used, but instead to predict Eulerian simulation properties. Thompson et al.~\shortcite{tompson2017accelerating} introduced a method exploiting CNNs to speed up existing solvers. They proposed a CNN-based architecture as a preconditioner for the pressure projection step. CNNs were also used to learn flow descriptors to visually enhance coarse smoke simulations using precomputed patches~\cite{chu2017data}. In a context where synthetic and plausible samples can be generated on-demand, using generative adversarial networks (GAN) seemed suitable for fluids. GANs were initially used as a super-resolution technique to enhance temporally coherent details of smoke simulations directly in image-space~\cite{xie2018tempogan}. Meanwhile, Kim et al.~\shortcite{kim2019deep} proposed another generative network for precomputing solution spaces for smoke and liquid flows. Super-resolution was later revisited for fluids, enabling an LSTM-based architecture to predict the pressure correction and generate physically plausible high-frequency details~\cite{werhahn2019multi}. Due to their sequential properties, LSTM-based layers were later used by~Wiewel et al.~\shortcite{wiewel2019latent} to predict changes of the pressure field for multiple subsequent timesteps. Another work on convolutional networks for Lagrangian fluids has been presented by Ummenhofer et al.~\shortcite{ummenhofer2019lagrangian}. They proposed a way to express particle neighborhoods using spatial convolutions as differentiable operators. In contrast with these methods, our network learns to encode differences between resolutions as opposed as to map fluid features using similarities.

Recently, Prant et al.~\shortcite{prantl2018generating} proposed a stacked neural network architecture allowing them to learn space-time deformations for interactive liquids. Although we share the same goal of precomputing simulation data of the liquid surface, our work focuses on the learning in a Lagrangian setup for up-resing purposes. Moreover, in comparison to the work of Prant et al.~\shortcite{prantl2018generating}, in which two stacked NNs are used to learn deformations, we propose a simpler and more efficient NN architecture by using a deformation coefficient directly in our objective function and by exploiting the particle neighborhoods to guide the convolution downsampling layers.

\begin{figure*}[t]
    \includegraphics[width=\textwidth]{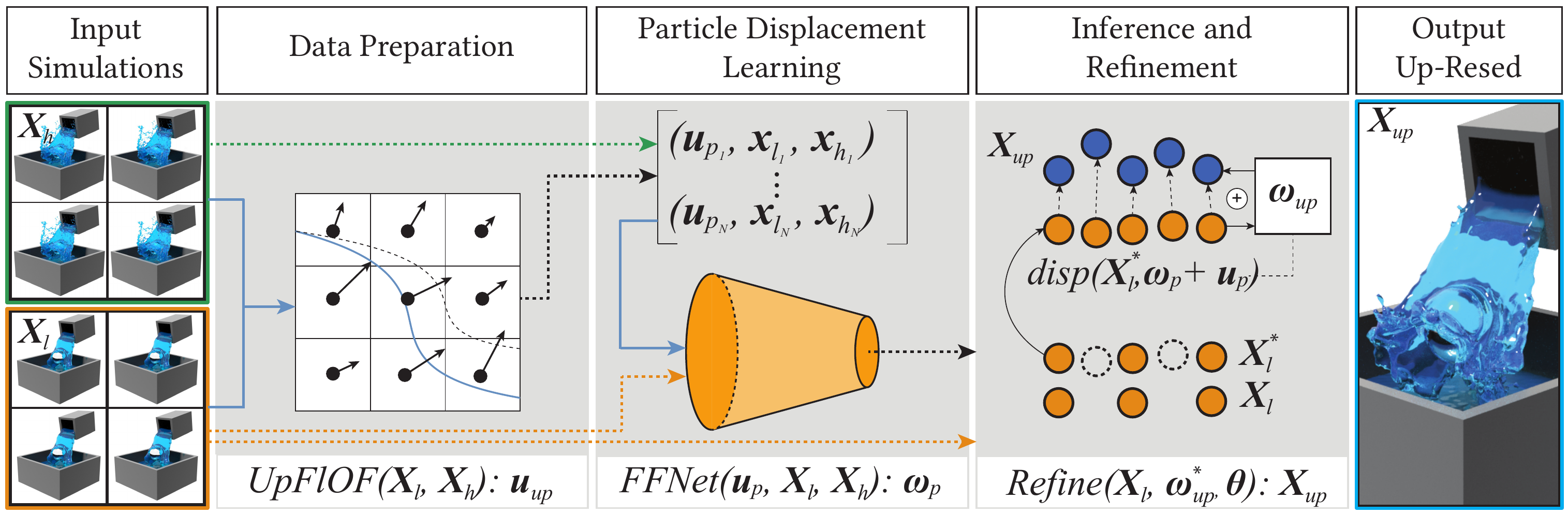}
    \caption{Overview of our up-resing pipeline. The input data used during the training phase is prepared using pairs of simulations at different resolutions (low and high), which is also used to generate an inter-resolution deformation field. Our network encodes particle displacements using differences between neighboring particles as a scene flow. These predicted displacements are then refined and applied directly to the particles to infer details.}
    \label{fig:overview}
\end{figure*}

\section{Method}
\label{sec:method_overview}
Given consecutive frames of an input animation of a liquid, our approach predicts Lagrangian deformations to infer high-resolution details and behavior. We propose an adapted scene flow learning architecture, called \textsl{FluidFlowNet} (\textsl{FFNet}), to determine Lagrangian motions on particle-based liquids. Our learning architecture is inspired from \textsl{FlowNet3D}~\cite{liu2019flownet3d}; it encodes displacements between two unstructured and unordered particle-based liquids simulated at different resolutions using solely their positions and velocities as input (\S~\ref{sec:network_architecture}). Our dataset is composed of over one thousand randomly generated pairs of liquid simulation scenarios in which each pair is composed of two simulations using the same initial parameters but at different resolutions: one simulation at a coarse resolution and the other at a high resolution (\S~\ref{sec:dataset}). Similarly to the work of Thuerey~\shortcite{thuerey2017interpolations}, we compute a mapping correspondence between the two liquids using an optical flow solve on a 4D volume. However, in our case, the resulting optical flow is used to capture the deformations between pairs of multi-resolution simulations guiding our loss function. We use the multi-resolution liquid samples as a training set to encode an inter-resolution deformation using our \textsl{FFNet} model. Finally, the inferred deformation field obtained from our learning pipeline is transposed into Lagrangian displacements and used as up-resing operators on the low-resolution particle-based liquid (\S~\ref{sec:post_processing}). With these displacements, we encode the missing details and behavior bounded from the input low-resolution simulation. In addition, a post-processing step refines the particles’ positions combining the low-resolution input velocities with the applied displacements while reducing the inference noise. Fig.~\ref{fig:overview} shows an overview of our up-resing pipeline for liquids. In this paper, we use bold symbols for vectors and capital symbols for matrices.


\paragraph{\textbf{Preparing Simulation Data}}
The dataset is built in three stages: (1)~generating many pairs of liquid animation scenarios (i.e., low-resolution and high-resolution) using different subsets of initial conditions, (2)~producing a deformation field using an optical flow solve on generated 4D volumes (used later in our loss function), and (3)~augmenting artificially the number of training samples using the deformation field to interpolate new variants of simulated scenarios.

\paragraph{\textbf{Learning Particle-Based Deformations}}
We use a neural network to learn and encode Lagrangian deformations between low- and high-resolution particle-based liquids. The proposed \textsl{FFNet} exploits the output of the optical flow solve to weigh each deformation between the input local features. The input features to our network are then expressed as the particle positions for the low- and high-resolution liquids.

\paragraph{\textbf{Inference and Refinement}}
Because we apply the deformations directly on particles, our approach firstly requires to upsample the particles close to the liquid interface before inference. Also, we generate our results using \textsl{Narrow Band FLIP} liquids to reduce the memory footprint since the deformations are mainly modeled according to surface particles. Finally, we deploy a refinement step at inference to account for generalization errors and to adjust the final particle motion according to the input velocity field.

\section{Inter-Resolution Liquid Interpolation}
\begin{figure}[h]
\centering
\begin{subfigure}[b]{0.19\linewidth}
    \centering
    \includegraphics[height=110px]{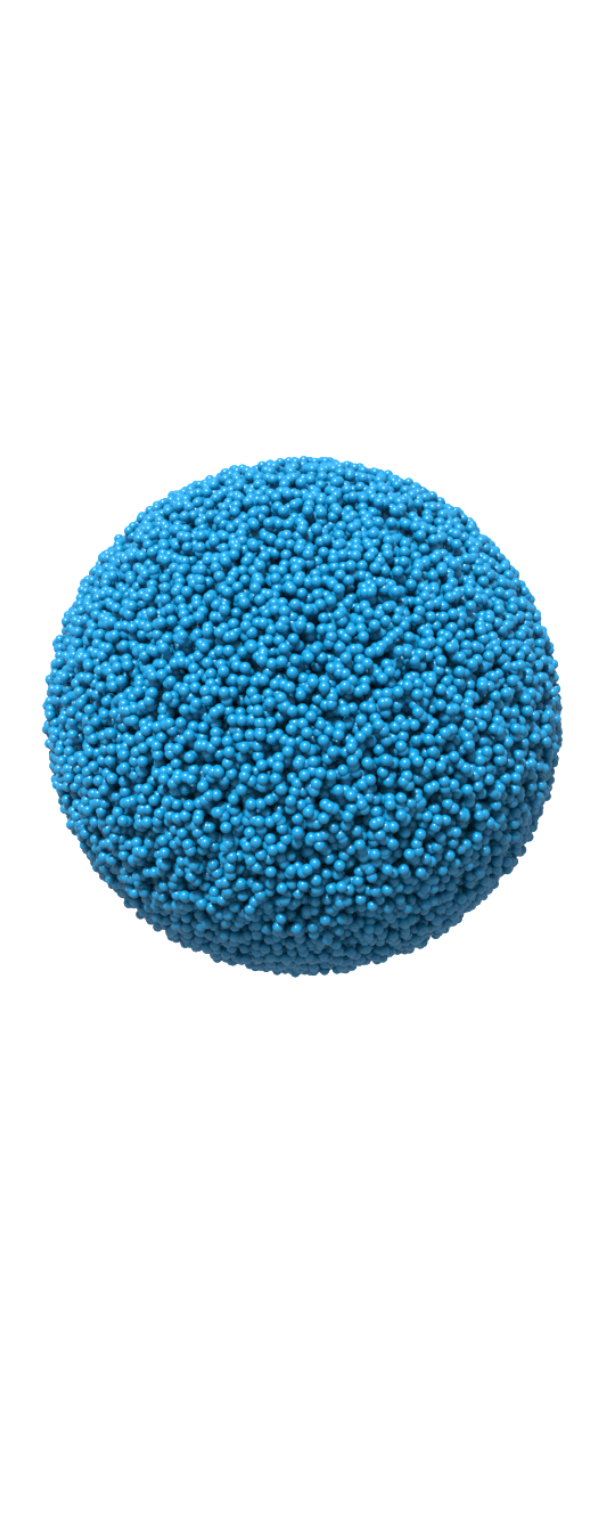}\\
    \caption{Input}
\end{subfigure}
\hfill
\begin{subfigure}[b]{0.59\linewidth}
    \centering
    \includegraphics[height=110px]{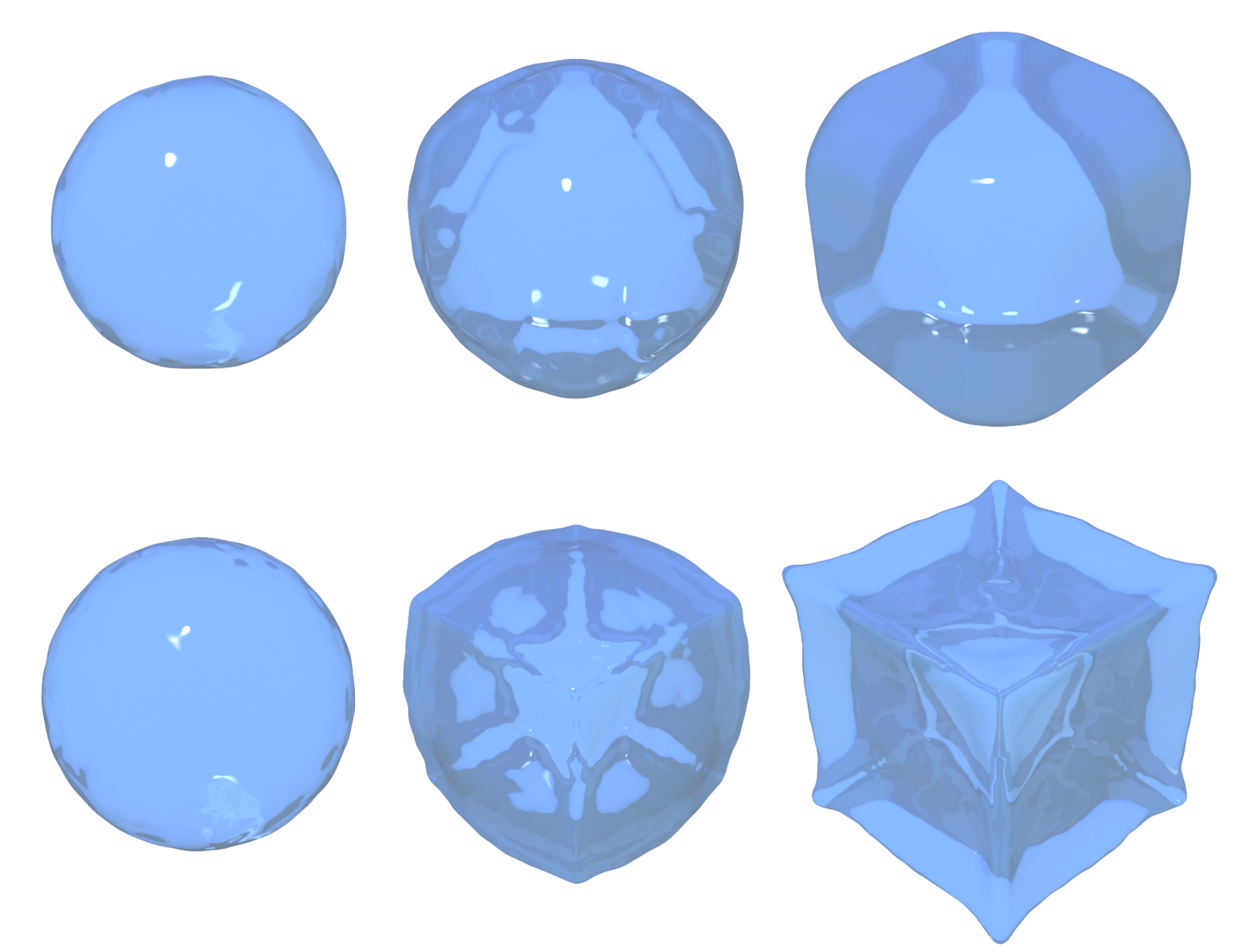}\\
    \caption{Interpolation}
\end{subfigure}
\begin{subfigure}[b]{0.19\linewidth}
    \centering
    \includegraphics[height=110px]{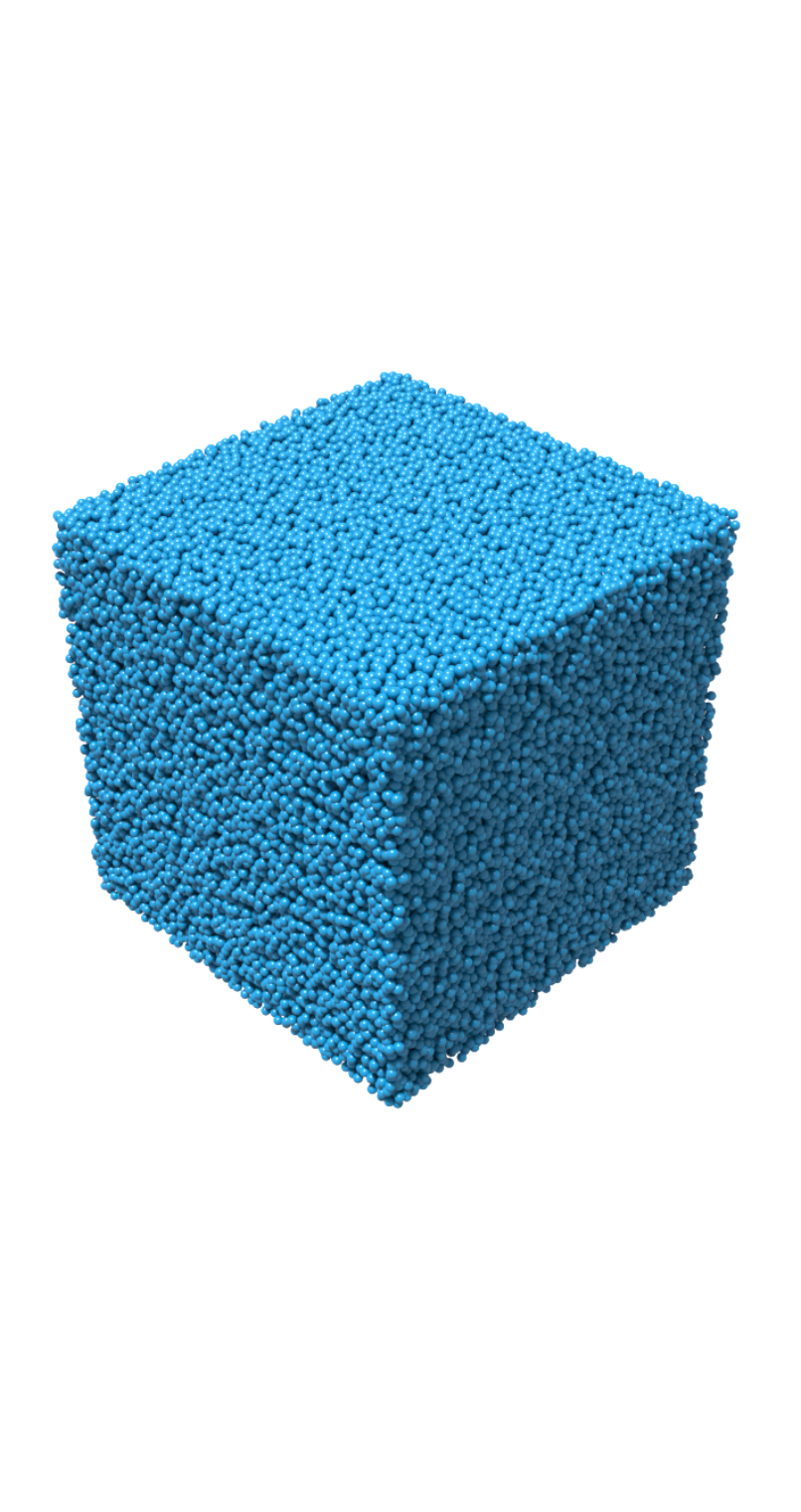}\\
    \caption{Target}
\end{subfigure}
\hfill
\caption{A simple case where a sphere is morphed at three steps~(b) into a cube: comparing our interpolation approach applied directly on particles~(bottom) with a state-of-the-art interpolation method~\cite{thuerey2017interpolations} applied on the corresponding SDF~(top).}
\label{fig:interpolation_comparison}
\end{figure}

In this section, we adapt an algorithm originally from Thuerey~\shortcite{thuerey2017interpolations} to interpolate between a low- and a high-resolution liquid. We use this interpolation approach in our proposed pipeline for two reasons: preparing the data for training with our neural network, and combining Eulerian deformations with the particle motions. Using an Eulerian inter-resolution deformation factor along with a particle-based deformation field ($\textbf{u}_p$ in Fig.~\ref{fig:overview}), our approach interpolates between resolutions directly on the particles. We are expressing these deformations in a Lagrangian manner to increase small-scale details and to offer a more intuitive control to infer deformations to particles. As highlighted in Fig.~\ref{fig:interpolation_comparison}, applying deformations directly on the particles allows a more faithful reproduction of fine details of the interpolation target (e.g., the sharp edges of the cube). In addition, we have observed that our approach gives a more realistic liquid appearance while preserving certain visual characteristics such as surface tension.

Our motivation in applying displacements directly to the particles is to better reflect the small-scale deformations while not being strictly bounded to the resolution of an underlying grid. An Eulerian deformation is quite successful in detecting smooth large-scale motions, but its inherent regularization prevents it from matching fine surface details at the cost of additional correction steps. Inspired by the method proposed by Thuerey~\shortcite{thuerey2017interpolations}, we use a 4D optical solve to interpolate between input scenarios using the same initial conditions, but at different resolutions (low- and high-resolution).

\subsection{Up-Resing Optical Solve}
\paragraph{\textbf{Inter-Resolution Optical Flow}}
We employ an approach similar to the work of Thuerey~\shortcite{thuerey2017interpolations}, that we briefly summarize in the following, as we propose a slightly altered form of it for learning purposes. Although our goal is also to compute a deformation field, we adapt Thuerey's method to interpolate a low-resolution input into one at a higher resolution, using the same initial conditions for both inputs. The generated deformation field is used in combination with the particles of each simulation (i.e., low- and high-resolution) as input features of the training set for our neural network (\textsl{FFNet}).

Given a pair of corresponding surfaces (i.e., generated SDF $\Phi$ from the input particles) using two distinct simulation resolutions (i.e., number of particles and grid resolution), the optical flow solve is expressed as a weighted sum of the energy terms to be minimized to compute the deformation field $\textbf{u}$:
\begin{equation}
\label{eq:min_energy_of}
\min_{\textbf{u}}~{E_{d}(\textbf{u})+\beta_S E_{\text{smooth}}(\textbf{u})+\beta_T E_{\text{Tikhonov}}(\textbf{u})},
\end{equation}
where the first term corresponds to the information related to the SDF (i.e., occupancy proportion values). The second and third terms are respectively the smoothness and Tikhonov regularizers. The smoothness term $E_{\text{smooth}}$ penalizes non-smooth solutions, and the Tikhonov term penalizes vectors with large magnitudes. The discretized minimization of Eq.~\ref{eq:min_energy_of} yields a system of linear equations $\textbf{A}_{\text{upof}}\textbf{u}=\textbf{b}$ where the first term $\textbf{A}_{\text{upof}}$ corresponds to the discretized energy terms given by:
\begin{equation}
\label{eq:discretized_energy_terms}
\textbf{A}_{\text{upof}}=\nabla{\Phi_{h}}^{T}\nabla{\Phi_{h}}+\beta_{S}\sum_j \textbf{L}_{j} + \beta_T\textbf{I},
\end{equation}
and where the terms are respectively the discrete spatial gradient squared, the smoothness (using the discrete Laplacian $\textbf{L}$), and the Tikhonov regularizers. Finally, we express the right-hand side $\textbf{b}$ of the linear system as $\left[\nabla{\Phi_{h}}\right]^T\Phi_{\delta}$ where $\Phi_{\delta}$ is the finite difference between the input surfaces $\Phi_{h}$ and $\Phi_{l}$. Solving the linear system for $\textbf{u}$ gives us the up-res deformation field $\textbf{u}_{up}$. Up to this point, our approach differs only from that of Thuerey~\shortcite{thuerey2017interpolations} by its preparation and application to the input data as we use it between simulation pairs of varying resolutions (see Algo.~\ref{algo:advection} for further details).

\begin{figure}[h]
\centering
\includegraphics[width=\linewidth]{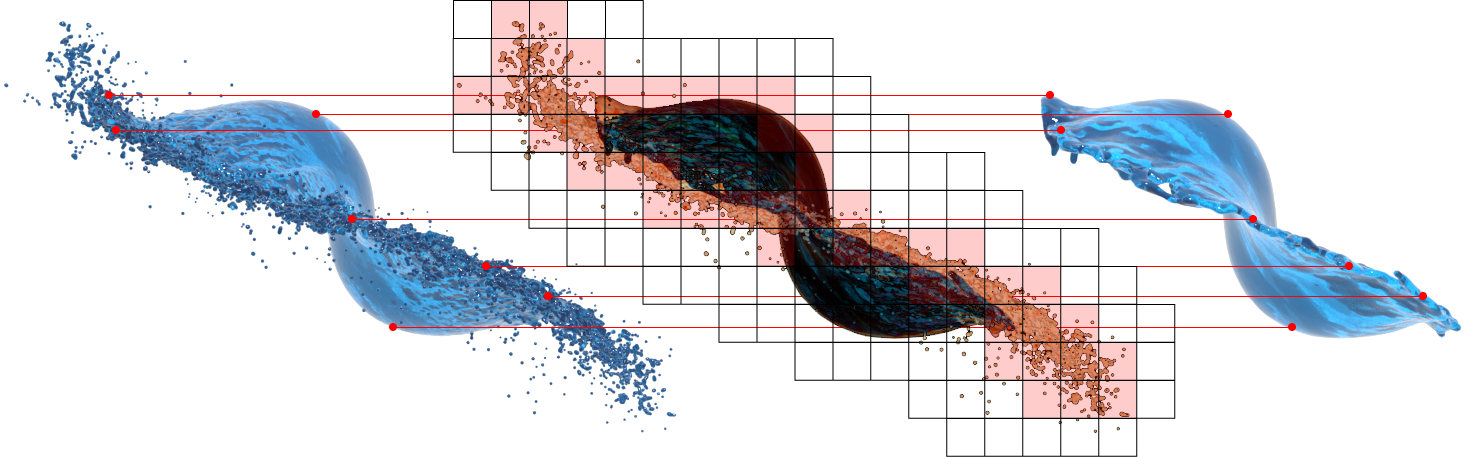}\\
\caption{As highlighted in this illustration, the key-event constraint term allows matching surface points between (right image) low- and (left image) high-resolution liquid surfaces. The complex cells are marked with a red background (middle image); the feature surface points are identified and connected using red points and lines.}
\label{fig:topological_changes}
\end{figure}

\paragraph{\textbf{Key-Event Alignment}}
As we focus on the differences between variable resolution inputs, we propose an additional and novel key-event alignment term directly within the optical flow solve to constrain the solution according to specific surface topological changes. The motivation is to align the deformations between inter-resolution inputs to capture and match certain key events in time. To do so, we define a soft constraint expressed as a penalty term~\textbf{D} detecting cells (using the grid of the SDF $\phi$) with topological changes. Since our method is already applying the optical solve on the SDF, we decided to use the \textit{complex cell tests}~\cite{wojtan2009deforming} to determine which cells are too complex to be represented with piecewise linear isosurfaces~\cite{varadhan2004topology}. Each cell complexity~$c_i$ (0 or 1) in~\textbf{D} is weighted using a proportional coefficient $d_{x_i\rightarrow x_k}$ based on its 4D Euclidean distance to the closest key surface point $\boldsymbol{x}_k$:
\begin{equation}
d_i=\frac{1}{d_{x_i\rightarrow x_k}}c_i,
\end{equation}
where $\boldsymbol{x}_k$ is the closest key surface point computed using the same feature point extraction method as used with point cloud registration. The feature points are selected using the local mean curvature approximation $\mu$ and compared to the domain distribution. We keep the feature points above a certain threshold with respect to their distribution $\mu+\alpha\rho$. The coefficient $\alpha$ is used to control the number of feature surface points preserved. The Euclidean distance $d_{x_i\rightarrow x_k}$ is computed using the first step of the Iterative Closest Point (ICP) algorithm which is used to determine the closest 4D key surface point $\boldsymbol{x}_k$ for each surface point at the position $\boldsymbol{x}_i$ (Fig.~\ref{fig:topological_changes}). Using the proportional coefficient $d_i$, we can then rewrite Eq.~\ref{eq:discretized_energy_terms} as follows:
\begin{equation}
\label{eq:discretized_of_align}
\textbf{A}_{\text{upof}}=\nabla{\Phi_{h}^{T}}\nabla{\Phi_{h}}+\beta_{S}\sum_j \textbf{L}_{j} + \beta_{T}\textbf{I} + \textbf{D}.
\end{equation}

\begin{figure}[h]
    \includegraphics[width=\linewidth]{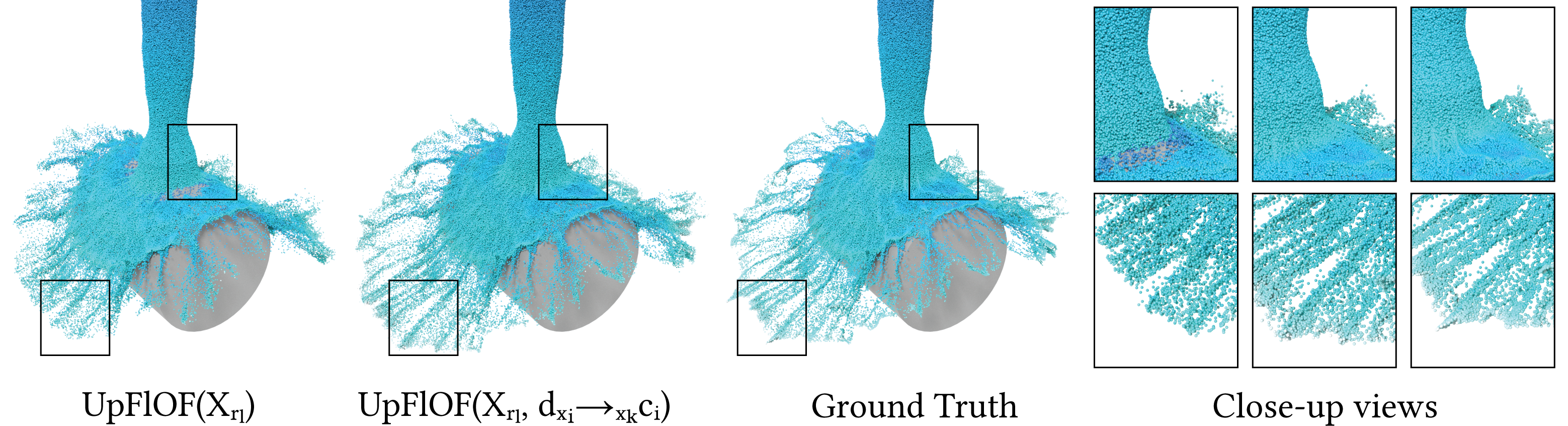}
    \caption{Comparing the deformation precision of $\textbf{u}_{up}$ when applied to a low-resolution surface using our key-event alignment term.}
    \label{fig:compare_key_alignment}
\end{figure}

As shown in Fig.~\ref{fig:compare_key_alignment}, small-scale artifacts arise when trying to interpolate between two liquids having major differences bounded by their resolution. By aligning the deformations $\textbf{u}_{up}$ using key-event feature surface points, we improve the fidelity (i.e., as close as possible to the high-resolution ground truth) of deformation on low-resolution input.

The motivation behind using the proportional coefficient $d_i$ with complex cells is to influence the deformation solution $\textbf{u}_{up}$ according to key cells (i.e., cells containing key points) defining a correspondence between two simulations. However, the discretization of a simulation may produce significant differences that can make it difficult to determine a match. This is the reason why we decided to opt for a soft constraint instead of a hard constraint because matching these key events was tricky to satisfy in some scenarios. 

\subsection{Dataset and Data Augmentation}
\label{sec:dataset}
Our dataset is composed of pairs of simulated \textsl{Fluid Implicit Particle (FLIP)} liquids using initial conditions from a parameter matrix~$\Theta$. As previously stated, each pair is composed of two simulated liquids: one at a low resolution and the other at a higher resolution. We adjusted the liquid resolution by changing the particle spacing (i.e., particle density per cell) and the underlying grid resolution. For all of the presented examples, we used a particle separation (ps) of $0.02$ and $0.005$, and a grid scaling factor (i.e., used to compute the cell size with the particle spacing) (gs) of $2.0$ and $1.2$, respectively for low- and high-resolution liquid samples. We define a pair of sample liquids as $P_i=\{ X_{l}, X_{h} \}_i$ where each sample $X_i$, regardless of its resolution, is expressed as a set of particle coordinates $\boldsymbol{x}_i$.

\begin{algorithm}[t]
\SetAlgoLined
\SetCommentSty{textrm}
\DontPrintSemicolon
\SetKwInOut{Input}{Input}
\Input{\begin{tabular}{ l }
$
\begin{rcases*}
    \text{o}_{\text{st}}\text{: obstacle shape type} (\text{6 different shapes})\\ 
    \text{x}_{o}\text{: obstacle position} ({x}_{o} \in \mathbb{R}^3)\\
    \text{x}_{em}\text{: emitter position} ({x}_{em} \in \mathbb{R}^3)\\
    \text{cd: container dimensions} (cd \in \mathbb{R}^3)
\end{rcases*} = \Theta
$
\end{tabular}}
\hrulefill\\
$P = \emptyset$\\
\For(\tcc*[f]{\textit{for each parameter set $\Theta_i$}}){$\forall \Theta_i \in \Theta$}{
    $X_{l} = \text{simulate}(\Theta_i, \text{ps}=0.02, \text{gs}=2.0)$ \tcc*{\textit{FLIP solver}}
    $X_{h} = \text{simulate}(\Theta_i, \text{ps}=0.005, \text{gs}=1.2)$ \tcc*{\textit{FLIP solver}}
    $P \cup \{ X_{l}, X_{h} \}$\;
}
$P^* = P$\\
\For(\tcc*[f]{\textit{for each pair of simulated liquids $P_i$}}){$\forall P_i \in P$}{
    $P_j = \text{randomize}(\{P_j | j \ne i, j \in 0 \le j \le |P| \})$\;
    $\{X_{l}, X_{h}\}_i, \{X_{l}, X_{h}\}_j = {P_i, P_j}$\;
    $X^*_{l} = \text{UpFlOF}(\{X_l\}_i, \{X_l\}_j, \Phi(\{X_l\}_i), \Phi(\{X_l\}_j), \alpha=0.5)$\;
    $X^*_{h} = \text{UpFlOF}(\{X_h\}_i, \{X_h\}_j, \Phi(\{X_h\}_i), \Phi(\{X_h\}_j), \alpha=0.5)$\;
    $P^* \cup \{ X^*_{l}, X^*_{h} \}$\;
}
\caption{Pseudo-code for generating and augmenting our training dataset.}
\label{algo:dataset_augmentation}
\end{algorithm}

\begin{figure}[h]
    \includegraphics[width=\linewidth]{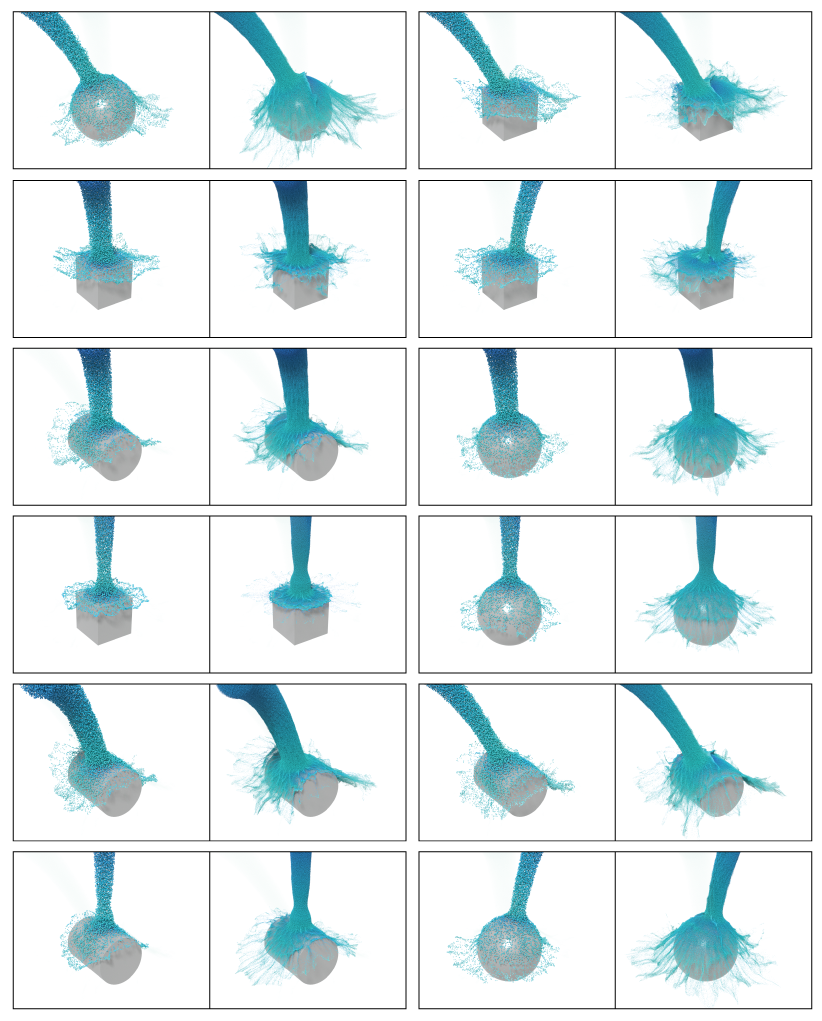}
    \caption{A small subset of our multi-resolution dataset used to train the proposed \textsl{FFNet} model. Each simulation used to generate our dataset is composed of a pair of liquids (each pair using the same initial conditions): a coarse low resolution and a detailed high resolution.}
    \label{fig:collide_dataset}
\end{figure}

The simulated liquids are generated using selected initial conditions taken from a matrix (as shown in Algo.~\ref{algo:dataset_augmentation}). We refer to each of these initial parameters using a subscript to our parameter matrix~$\Theta \in \mathbb{R}^{n\times m}$ where $n$ is the number of parameters, and $m$ the number of initial values for each parameter (e.g., obstacle shape type $\text{o}_{\text{st}}$,  obstacle position $\text{x}_{o}$, emitter position $\text{x}_{em}$, container dimensions cd). As an example, the notation $\Theta_{\text{x}_{em}}$ would be for an emitter position of a sample simulation. The resolution parameters, particle spacing $ps$ and grid scale $gs$, are not included in the matrix because they are fixed for all pairs of simulated liquids. For our dataset purpose, the emitter velocity is computed with respect to the emitter position in order to be oriented toward an existing obstacle and/or container. The types of shapes $o_{\text{st}}$ can be either used as static boundaries or as a liquid initial shape. We present in Fig.~\ref{fig:collide_dataset} a small subset of generated samples for the training set.

\paragraph{\textbf{Data Augmentation}}
As shown in the second half of Algo.~\ref{algo:dataset_augmentation}, we use data augmentation on the simulated liquids to artificially grow the number of samples within our dataset. In this regard, we use the original FlOF algorithm as proposed by Thuerey~\shortcite{thuerey2017interpolations} to interpolate liquids between precomputed initial conditions. We first randomly select each simulation pair $P_i$ with a pair $P_j$ to generate a new pair of liquids. Then, we extract for each pair their associated particle coordinates $X$ as inputs to the interpolation method. The interpolated liquid is obtained by computing a bidirectional optical flow $\boldsymbol{u}_{\omega}$ applied on the surface $\Phi(X)$. We use $\alpha=0.5$ to generate an in-between simulation for each interpolated liquid (see \textsl{UpFlOF} method in Algo.~\ref{algo:advection}).

Even though the resulting liquids generated by interpolation are inevitably not physically accurate, they still provide significant additional data points allowing us to improve the generalization rate when it comes to predicting complex high-resolution behaviors.

\section{Scene Flow Learning for Lagrangian Deformation on Particles}
In this section, we will further detail the network architecture of our proposed approach (\S~\ref{sec:network_architecture}) and our neighborhood-based loss function for particle-based liquids (\S~\ref{sec:loss_function}).

\begin{figure}[b]
\centering
\includegraphics[width=\linewidth]{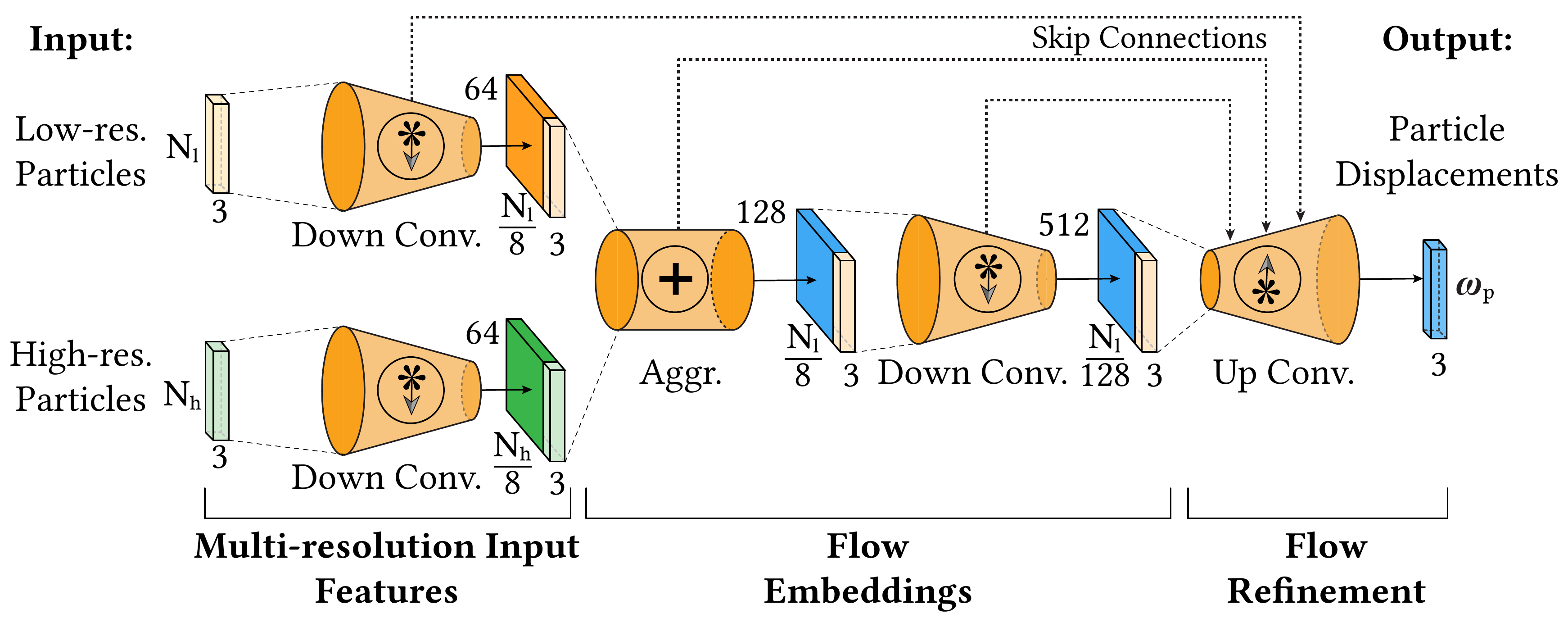}
\caption{The proposed network architecture. Given a pair of input liquid simulations, the particle deformation network focuses on encoding the particle displacement between low- and high-resolution particle-based simulations guided by the outputs from the UpFlOF approach. The dimensions at each convolution layer is expressed as a portion of the input resolution (N: number of particles) by 3 (number of components of the input feature). The number of local features grows as we progress down in the downsampling convolutions (e.g., 32, 64, 128, and so on).}
\label{fig:networks_architecture}
\end{figure}

\subsection{Network Architecture}
\label{sec:network_architecture}
As previously mentioned in \S~\ref{sec:method_overview}, our proposed learning architecture (Fig.~\ref{fig:networks_architecture}), which we call \textsl{FFNet} is an adaptation of the \textsl{FlowNet3D} method~\cite{liu2019flownet3d} in order to successfully capture the inherent and latent liquid properties. By giving additional cues of the underlying liquids, our learning approach converges faster to a more precise displacement solution.

Inspired by \textsl{FlowNet3D}, our network architecture is divided into three main modules: multi-resolution input features, flow embedding, and flow refinement. We will describe how we adapted each module in the following.

\paragraph{\textbf{Multi-Resolution Input Features}}

In this first part of our learning pipeline, we use the local neighborhoods to estimate the convolution operators, as proposed with the \textsl{PointNet++} architecture~\cite{qi2017pointnet++}. Since a traditional convolution does not work with unstructured data points such as a particle set, using a downsampling approach based on the neighborhood provides a suitable way to encode these local features.

\begin{figure}[h]
    \includegraphics[width=\linewidth]{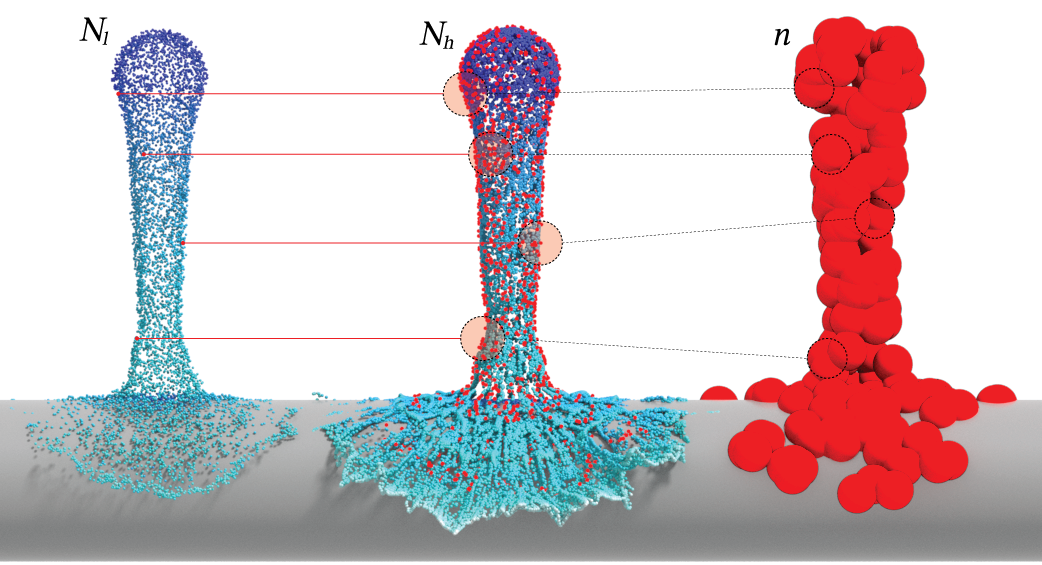}
    \caption{At each iteration for our network, a correspondence map is generated between the input pairs of multi-resolution simulations $X_l$ and $X_h$ consisting of respectively $N_l$ and $N_h$ particles. These $n$ correspondences are expressed as representative neighborhood for each level of convolution.}
    \label{fig:sampling_regions}
\end{figure}

As shown in Fig.~\ref{fig:sampling_regions}, a convolution layer downsamples an input liquid with $N$ particles into $n$ neighborhoods. Each particle $p_i = \{\boldsymbol{x}_i, \boldsymbol{f}_i\}$, where $\boldsymbol{x}_i$ is the $\mathbb{R}^3$ coordinates of particle $i$ and $\boldsymbol{f}_i$ its associated local features generated at each layer of our network. Using the SPH neighborhood computation~\cite{becker2007weakly}, we consider the contributions of each particle $p_i$ in its neighborhood using weighted averages. As opposed to computing each weighted average at the position of a particle, we compute the contributions at the center of each downsampled neighborhood $n_j$. The weighted local features $f^*_i$ for a downsampled neighborhood are then computed as follows:
\begin{equation}
f^*_i(x)=f_i|x-\overline{x}_{n_j}|,
\end{equation}
where $x$ is the center coordinates of neighborhood $n_j$, and $\overline{x}_{n_j}$ is the weighted average of the coordinates of particles in that neighborhood:
\begin{equation}
\boldsymbol{\overline{x}}_{n_j} = \sum_{i \in n_j} w_i \boldsymbol{x}_i.
\end{equation}
The weights for each neighborhood $n_j$ are computed using the same smooth kernel function introduced by Zhu and Bridson~\shortcite{zhu2005animating}:
\begin{equation}\label{eq:kernel_function}
w_i(x)=\frac{k(|\boldsymbol{x}-\boldsymbol{x}_i|)R^{-1}}{\sum_j k(|\boldsymbol{x}-\boldsymbol{x}_j|)R^{-1}},
\end{equation}
where $k(s)=\max(0,(1-s^2)^3)$ for a smooth transition between neighbor contributions, and radius $R$ is equal to twice the particle separation used to generate the input simulation. We express the weighted features as the importance level of each local feature during the convolution phases.

We have noticed throughout experiments that using the weighted average to estimate the local features for each downsampled neighborhood $n_j$ improved the recall when predicting large-scale flows. Finally, we use an element-wise max pooling operator $\textbf{MAX}$ on a nonlinear approximation using a multi-layer perceptron (MLP) on the weighted $f^*_i(x)$ as follows:
\begin{equation}
f^*_{n_j}=\underset{i | \{\|\boldsymbol{x}_i-\boldsymbol{x^{\prime}}_{j}\|\le R\}}{\textbf{MAX}}\{ h(\boldsymbol{f}^*_i, \boldsymbol{x}_i-\boldsymbol{x}^{\prime}_{j})\},
\end{equation}
where $h$ is the nonlinear MLP function, and $R$ the same radius as used in Eq.~\ref{eq:kernel_function}.

\begin{figure*}[t]
    \includegraphics[width=\textwidth]{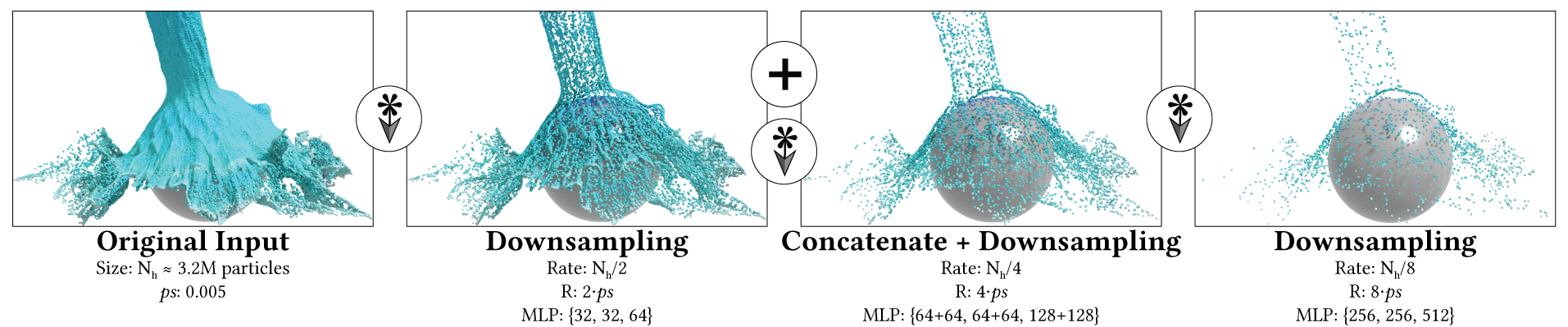}
    \caption{Example of a cascade of downsampling convolution operations applied on a high-resolution particle-based liquid.}
    \label{fig:conv_downsampling}
\end{figure*}

\paragraph{\textbf{Fluid Flow Embedding}}
Once the weighted local features are computed for an input pair (i.e., for both low and high resolution), the two local features $f^*_{n_j}$ and $g^*_{n_j}$, respectively generated from the low-resolution input $X_l$ and the high-resolution input $X_h$, are combined before applying convolution to training samples.

Similarly to creating the multi-resolution input features, we use the same number of neighborhoods to describe each particle-based input liquid pair, even if they are discretized differently. That way, our network can define the flow embeddings using the same neighborhood center coordinates $\boldsymbol{x}_{n_j}$ for a simulation pair alongside the weighted local features and the original particle coordinates. Again, we use the same weighted function $w_i(x)$ as input to the nonlinear MLP function $h$ to aggregate each neighbor contribution prior to being max pooled. We express each particle embedding as follows: 
\begin{equation}
e_i=\underset{i | \{\|\boldsymbol{x}_i-\boldsymbol{x}^{\prime}_{j}\|\le R\}}{\textbf{MAX}} \{ h(\boldsymbol{f}^*_{n_j} \oplus \boldsymbol{g}^*_{n_j}, \boldsymbol{x}_{l}^{i}-\boldsymbol{x}_{h}^{i}) \},
\end{equation}
where $\boldsymbol{f}^*_{n_j}$ aggregates the input features of the downsampled neighborhood $n_j$ from the low-resolution particle-based liquid, and $\boldsymbol{g}^*_{n_j}$ the input features of the downsampled neighborhood $n_j$ from the high-resolution one. We perform a few additional convolutions to spatially smooth out the corresponding features with respect to each particle of each neighborhood.

\paragraph{\textbf{Flow Refinement}}
Finally, as the last step of the network, we focus the learning on propagating the embedded features $e_i$ from the downsampled neighborhood $n_j$ to the original low-resolution input particles $N_{l}$. We use upsampling convolution operations to learn how to project weighted features back to the input particle coordinates. We have noticed throughout experiments that propagating the weighted features from the embedding layer improved the ability of the network to recover nonlinear features specific to fluids, such as vorticity.

We show the precision of our refinement step on downsampled features (Fig.~\ref{fig:conv_downsampling}) as opposed to solely using a symmetric MLP function on these. A last and dense regression layer is used after the upsampling convolutions to project back the scene flow prediction in $\mathbb{R}^3$ (i.e., without using an activation function). Also note that as shown in Fig.~\ref{fig:networks_architecture}, we use skip connections between downsampling and upsampling layers to infer multi-scale feature learning.

\subsection{Deformation-Aware Loss Function}
\label{sec:loss_function}

As explained previously, we are using pairs of particle-based simulations as input samples. Our training loss on these samples is evolving throughout epochs to capture the differences in displacements between simulations using the same initial condition parameters, but at different discretizations. We express our loss function as $L_1$ metrics to encode absolute differences between the low-resolution and high-resolution samples:
\begin{equation}
\label{eq:loss_function}
L_{up}=\frac{1}{n_j}\sum_{i}^{n_j}{\|\boldsymbol{\omega}_i-\boldsymbol{\omega}_i^* \|_1 + \lambda\| \boldsymbol{\omega}_i^{\leftarrow}-\boldsymbol{\omega}_i \|_1},
\end{equation}
where $\boldsymbol{\omega}_i$ is the predicted displacement and $\boldsymbol{\omega}_i^*$ is the ground truth. Interestingly, the cycle-consistency regularization term $\| \boldsymbol{\omega}_i^{\leftarrow}-\boldsymbol{\omega}_i \|_1$ (in Eq.~\ref{eq:loss_function}) acts as a penalization constraint to enforce the bidirectionality of the resulting scene flow for displacements. In other words, this term enforces that the forward flow and the backflow flow (i.e., displacement between the predicted particle coordinates and the displaced one $\boldsymbol{\omega}_i^{\leftarrow}$ using $\boldsymbol{\omega}_i^*$) closely match each other. Also, we have noticed empirically that adding an adaptive weight $\lambda_{n_j}$ to adjust the neighborhood $n_j$ contributions (as opposed to using $\lambda=1$ for every neighborhood) has improved significantly the convergence rate throughout iterations:
\begin{equation}
L_{up}=\frac{1}{n_j}\sum_{i}^{n_j}{\|\boldsymbol{\omega}_i-\boldsymbol{\omega}_i^* \|_1 + \boldsymbol{\lambda}_{n_j}\| \boldsymbol{\omega}_i^{\leftarrow}-\boldsymbol{\omega}_i \|_1},
\end{equation}
where the adaptive weight $\lambda_{n_j}$ is computed using the normalized magnitude of the deformation initially computed with the \textsl{UpFlOF} algorithm. From our analysis comparing different training experiments, we have observed that using such an adaptive weight was allowing emphasis on encoding features where the main differences between discretized simulations were occurring within neighborhoods. Throughout the experiments, we also observed that the proposed adaptive weight improved (in some of the results presented) the reconstruction of small complex details, as we were able to train through more iterations without overfitting the resulting generalization model.


\section{Deformation Inference and Refinement}
\label{sec:post_processing}
In the following sections, we dive into the final steps of applying the predicted displacements onto the input particle-based liquid. First, we apply the displacements taking into account the existing velocities of the input particles. Then, a refinement step is performed to reduce the displacement noise generated by the inference.



\subsection{Applying the Displacement on Particles}
\label{sec:particle_displacement}
In this section, we explain the steps performed by our approach to combine the input velocity with the predicted Lagrangian displacements $\boldsymbol{\omega}_p$. Firstly, we upsample the input narrow band $X$ of width $d_b$ to ensure that the density of the upsampled particle set $\boldsymbol{X}^{'}$ is high enough to capture the small-scale deformations generated by our resulting deformation field. Once the predicted displacements are generated, we transfer them to a velocity field~$\boldsymbol{u}_{\omega}$. 

The cell size of $\boldsymbol{u}_{\omega}$ is adjusted with respect to the number of particles per cell preventing undesirable gaps in the liquid during motion, as suggest by Zhu and Bridson~\shortcite{zhu2005animating}. We resample the velocity field in a MAC grid $\boldsymbol{u}_{\text{MAC}}$ that will then be updated by extrapolation within a distance $d_{\text{MAC}}$ outside the SDF $\boldsymbol{\Phi}_{l}^{\prime}$ in order to fully cover the simulation domain of the upsampled input particles~$X^{\prime}$. Using the updated $\boldsymbol{u}_{\text{MAC}}$, we can now update our displacements to compensate for the actual input motion. That way, our approach can infer Lagrangian displacement regardless of the input motion. The MAC grid $\boldsymbol{u}_{\text{MAC}}^{\prime}$ is weighted according to the resampled velocity field $\boldsymbol{\hat{u}}_{\omega}$ and added to the current velocity field $\boldsymbol{\hat{u}}_{\omega}$. Finally, we advect the particles in a grid (as with FLIP) using $\boldsymbol{\hat{u}}_{\omega}^{\prime}$ on the upsampled particles $\boldsymbol{X}^{\prime}$ at each time step. Although our approach requires switching back and forth between Eulerian and Lagrangian deformations, we have empirically noticed that our model learns better to preserve fine details on the surface, as opposed to learning to deform these solely in an Eulerian manner. An overview summarizing these steps is presented with Algo.~\ref{algo:advection} (with isInference() equals to \textsl{true}).

\begin{algorithm}[t]
\SetAlgoLined
\SetCommentSty{textrm}
\DontPrintSemicolon
\SetKwInOut{Input}{Input}
\Input{\begin{tabular}{ l }
$\boldsymbol{X}$: set of particles\\
$\boldsymbol{u}_{\textnormal{MAC}}$: low-res. MAC velocity grid\\
$d_{\text{MAC}}$: extrapolation distance (2 cells)\\
$d_b$: width of narrow band (2-3 cells)\\
$r_h$: grid resolution larger than the input
\end{tabular}}

$\boldsymbol{\Phi}_{l}=\text{computeSDF}(\boldsymbol{X})$\\
$\boldsymbol{X}^{\prime}=\text{resampleNarrowBand}(\boldsymbol{X}, d_b)$\\
\uIf(\tcc*[f]{\textit{for predicting displacements}}){\textnormal{isInference()}}
{
    $\boldsymbol{\omega}_{p}=\text{predict}(X)$ \tcc*[f]{\textit{see Fig.~\ref{fig:networks_architecture}}}\\   $\boldsymbol{u}_{\omega}=\text{transferToGrid}(\boldsymbol{\omega}_{p})$\\
    $\boldsymbol{\Phi}_{l}^{\prime}=\boldsymbol{\Phi}_{l}$
}
\Else(\tcc*[f]{\textit{for data augmentation}})
{
    $\boldsymbol{\Phi}_{l}^{\prime}=\text{upscaleInterpolate}(\boldsymbol{\Phi}_{l}, r_h)$\\
    $\boldsymbol{u}_{\omega}=\text{FlOF}(\boldsymbol{\Phi}_{l}^{\prime}, \boldsymbol{\Phi}_{h}, \alpha=1)$ \tcc*[f]{\textit{see Algo.~\ref{algo:dataset_augmentation} in~\cite{thuerey2017interpolations}}}
}
$\boldsymbol{\hat{u}}_{\omega}=\text{resampleOFtoMAC}(\boldsymbol{u}_{\omega}, \boldsymbol{u}_{\text{MAC}})$\\
$\boldsymbol{u}_{\text{MAC}}=\text{extrapolate}(\boldsymbol{u}_{\textnormal{MAC}}, \boldsymbol{\Phi}_{l}^{\prime}, d_{MAC})$\\
$\boldsymbol{u}_{\text{MAC}}^{\prime}=\boldsymbol{\hat{u}}_{\omega}\boldsymbol{u}_{\text{MAC}}$\\
$\boldsymbol{\hat{u}}_{\omega}^{\prime}=\boldsymbol{\hat{u}}_{\omega}+\boldsymbol{u}_{\text{MAC}}^{\prime}$\tcc*[f]{\textit{inject input motion}}
$\boldsymbol{X}_{\text{advect}}=\text{advectInGrid}(\boldsymbol{X}^{\prime}, \boldsymbol{\hat{u}}_{\omega}^{\prime}, t)$ \tcc*[f]{$t=t+\triangle{t}$}
\caption{Pseudo-code for UpFlOF and inferred semi-Lagrangian advection.}
\label{algo:advection}
\end{algorithm}

\subsection{Upsampling and Reducing Surface Noise}

We have noticed during our experiments that surface noise was introduced at the inference step when compared to the ground truth. In order to validate the source of this noise, we have investigated on the resulting deformations before combining them with the input motion (i.e., exclusively based on $\boldsymbol{\omega}_p$). Similar to regression methods with point clouds (such as used with scene flow), this regression noise was appearing in the downsampling operations during the learning phase. As suggested by Liu et al.~\shortcite{liu2019flownet3d}, performing multiple passes of inference on randomized resampled samples using average predictions improved the results, but at the cost of losing some of the small-scale liquid behavior mostly perceptible during motion.

Since our goal is to focus the deformation close to the surface, performing a regression on the whole particle-based fluid seemed like an inadequate solution. As a matter of fact, the resulting noise came mostly from particles emerging from deep below the surface. We then determined that a resampling approach as used in \textsl{Narrow Band FLIP}~\cite{ferstl2016narrow} would be more suitable to convey where to resample in order to prevent displacing particles deep in the liquid (as opposed to randomly resampling). We ended up performing multiple passes of inference using varying depths $d$ defining the particle band thickness.

\begin{figure}[h]
    \includegraphics[width=\linewidth]{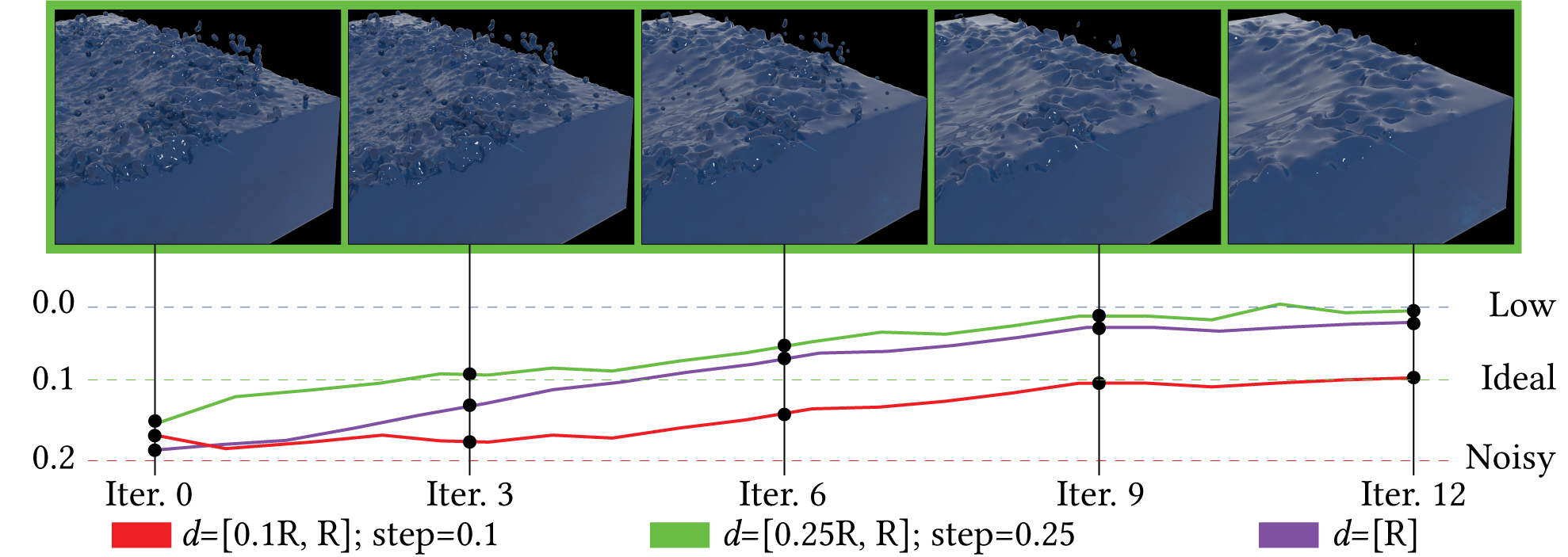}
    \caption{We compare the influence of varying upsampling distances $d$ over the surface noise (expressed as an error on the vertical axis) generated throughout the inference iterations. The images corresponding to the evolving level of noise (i.e., from iteration 1 to 12) of the green curve is presented on the top of the chart.}
    \label{fig:inference_noise}
\end{figure}

As shown in Fig.~\ref{fig:inference_noise}, we came up with a fair tradeoff throughout several experiments. For our purpose, this tradeoff is purely qualitative and might differ for other types of simulations. An error (between the displacement and the ground truth) smaller than $0.1$ is barely noticeable (especially in motion). Also, as exposed in Fig.~\ref{fig:inference_noise} (rightmost image generated after 12 iterations), performing more than 6 inference iterations is diluting the high-resolution features and ultimately converging back to the low-resolution input (i.e., fading out the displacements). The final displacement obtained is then averaged over all the iterations.

\section{Results}

In the following, we discuss the training experiments and expose the details of the corresponding datasets. We also demonstrate on multiple examples the capabilities of our approach to increase the apparent resolution using solely coarse particle-based simulations as input. We refer the reader to the supplemental video for the corresponding animations.

\subsection{Training and Inference}

\paragraph{\textbf{Datasets and Augmentation}}
\label{sec:datasets}
The simulations in our dataset have been generated using the Bifröst\textregistered~fluid solver for Maya\textregistered. The simulations for both resolutions are computed using varying parameters as exposed in Algo.~\ref{algo:dataset_augmentation}. We have divided the datasets into three categories referred to as \textsl{Colliding}, \textsl{Shape}, and \textsl{Container} (see Table~\ref{table:datasets} for further details). These datasets are covering many simple and specific simulation cases to improve inference generalization. The \textsl{Colliding} dataset contains simulations in which an emission source hits a single collision shape (Fig.~\ref{fig:collide_dataset}). The \textsl{Shape} dataset contains simulations in which a single liquid volume (initialized with different shapes) falls into an empty container. Lastly, the \textsl{Container} dataset contains simulations in which an emission source pours into a liquid container. For all these datasets, we define the breadth of simulation scenarios used for training by enumerating all combinations from a fixed set of obstacle shapes $o_{st}$, obstacle positions $\boldsymbol{x}_{o}$, position of emission sources $\boldsymbol{x}_{em}$, and dimensions of container cd. The shapes were determined to cover different shapes including sharp edges (cube), curved faces (sphere), and a mixture of both (cylinder).


\begin{table}[H]
\centering
\begin{tabular}{|l||c|c|c|c|}
    \hline
    \multirow{2}{*}{\textbf{Dataset}} & \multicolumn{2}{|c|}{\textbf{Simul. res. (low/high)}} & \textbf{Aug.} & \textbf{\# simul.}\\\cline{2-3}
      & \textbf{\# particles} & \textbf{Veloc. grid} & \textbf{factor} & \textbf{samples}\\[0.5ex]
    \hline\hline
    \textsl{Colliding} & 300k/2.4M & $96^3$/$192^3$ & $\times2$ & 432\\
    \hline
    \textsl{Shape} & 150k/1.5M & $96^3$/$192^3$ & $\times2$ & 432\\
    \hline
    \textsl{Container} & 400k/3.2M & $128^3$/$256^3$ & $\times12$ & 432\\
    \hline
\end{tabular}
\caption{Datasets of simulation pairs used as training and validation sets. We also present the augmentation factor (aug. factor) used to grow our datasets prior to training.}
\label{table:datasets}
\end{table}

In order to artificially grow each of these datasets for training, we interpolate each simulation pair using Thuerey’s method~\cite{thuerey2017interpolations} on shape's SDF, shape position, and emission position. On both \textsl{Colliding} and \textsl{Shape} datasets, we interpolated with weight $\alpha=0.5$, giving us an augmentation factor of $\times2$. We used more interpolation weights on the \textsl{Container} dataset since it originally consisted of fewer simulations. Each simulation pair of the \textsl{Container} dataset was interpolated at $\alpha \in \{0.25, 0.5, 0.75\}$ giving us an augmentation factor of $\times12$. Finally, we used a data split of 90\%-10\% respectively for training and validation sets. The test set was composed of new and significantly different simulation setups as enumerated in Table~\ref{table:parameters}.

\paragraph{\textbf{Performance Analysis}}

We have tested our approach on a variety of simulation setups to validate its robustness when generalizing within unknown initial conditions. As shown in Table~\ref{table:parameters}, the input simulations we used to test our learning model are of a fairly coarse resolution. The simulation time to generate the coarse simulation in Maya\textregistered~is presented since that simulation data is used as input to our network for inference. As previously mentioned, since we are mostly interested in surface details, we opted for the \textsl{Narrow Band FLIP} method to generate the particle-based liquids used for both training and testing inference. The computation times presented are expressed in seconds and computed at each frame of simulation and inference. The computation times of our \textsl{FFNet} network at inference are also presented. The interesting aspect when looking at these is that they are only slightly influenced by the input complexity. Another benefit of using the neighborhood convolutions to downsample multi-resolution particle-based simulations is that the evaluation times at inference turns almost constant and decoupled from the input resolution as we progress down the convolution layers.

\begin{table}[H]
\centering
\begin{tabular}{|l||c|c|c|c|}
    \hline
    \multirow{2}{*}{\textbf{Example}} & \multirow{2}{*}{\textbf{\# part.}} & \textbf{Grid} & \multirow{2}{*}{\textbf{Sim.}} & \textbf{\textsl{FFNet}}\\
    &  & \textbf{res.} & & \textbf{eval.}\\[0.5ex]
    \hline\hline
    \textsl{Pouring}: Fig.~\ref{fig:teaser} & 500k & $128^3$ & 0.724 & 0.071 \\
    \hline
    \textsl{Collision}: Fig.~\ref{fig:topological_changes} & 320k & $96^3$ & 0.542 & 0.061 \\
    \hline
    \textsl{Cylinder}: Fig.~\ref{fig:compare_key_alignment} & 290k & $96^3$ & 0.497 & 0.055 \\
    \hline
    \textsl{Stirring}: Fig.~\ref{fig:stiring_bowls} & 250k & $128^3$ & 0.482 & 0.058 \\
    \hline
    \textsl{Multi-stage}: Fig.~\ref{fig:multi_stage_streams} & 1.1M & $192^3$ & 1.249 & 0.112 \\
    \hline
    \textsl{Multi-streams}: Fig.~\ref{fig:acm_streams} & 900k & $192^3$ & 1.045 & 0.098 \\
    \hline
\end{tabular}
\caption{Statistics for the presented examples generated using our up-resing neural network \textsl{FFNet}. The discretization details (i.e., number of particles and velocity grid resolution) are shown for each example and the computation times per frame are expressed in seconds.}
\label{table:parameters}
\end{table}

The training was performed for 300 epochs on 1296 pairs of simulations, taking 8 hours on average for each epoch computed on two nodes of four NVIDIA\textregistered~GeForce\textregistered~RTX 2080 Ti. The \textsl{FFNet} architecture is composed of three downsampling convolution layers (detailed in Fig.~\ref{fig:conv_downsampling}), one embedding layer, and three upsampling convolution layers. An aggregation operation is performed after the first downsampling convolution layer to combine both input resolutions. Lastly, a linear flow regression layer is added at the end of the learning pipeline to output predicted particle displacements in $\mathbb{R}^3$. We also use skip connections at each convolution level to concatenate the outputted local features between downsampling and upsampling layers. Finally, the MLP function $h$ is activated by rectified linear units (ReLU) preceded by batch normalizations for both downsampling and upsampling convolution layers.

We chose a few baselines to compare the efficiency of our approach at the evaluation stage. In fairness and validity, we constrained the comparison to exclusively point-based learning methods. The evaluation metrics selected to compare the validity of these methods with respect to ours in the application context are the estimated position error (EPE) and the accuracy of the predicted displacement (Flow accuracy) when compared to ground truth. The EPE is evaluated as the average $L_2$ distance between the predicted and the ground truth displacement vectors. Since the number of particles of the ground truth might differ from the up-resed one generated with our approach (i.e., upsampled $X_l^*$ of the $X_l$ coarse input), we use the closest particle to match each particle of the reference particle-based liquid. The flow accuracy measures the proportion of predicted displacements (using a small error margin $\epsilon=0.001$ with respect to the scene scale) that are below a certain threshold (we used 0.1).

\begin{table}[h]
\centering
\begin{tabular}{|l||c|c|c|}
    \hline
    \multirow{2}{*}{\textbf{Method}} & \multirow{2}{*}{\textbf{EPE}} & \textbf{Flow} & \textbf{Conv.}\\
     & & \textbf{accuracy} & \textbf{time}\\[0.5ex]
    \hline\hline
    \textsl{PointNet}~\cite{qi2017pointnet} & 0.45 & 26.11\% & 14.4\\
    \hline
    \textsl{PointNet++}~\cite{qi2017pointnet++} & 0.44 & 29.84\% & 13.6\\
    \hline
    \textsl{FlowNet3D}~\cite{liu2019flownet3d} & 0.37 & 52.27\% & 10.3\\
    \hline
    \textsl{FFNet} (ours) & \textbf{0.23} & \textbf{63.71\%} & \textbf{8.1}\\
    \hline
\end{tabular}
\caption{Flow estimation results on the \textsl{Colliding} dataset.}
\label{table:compare_star}
\end{table}

\begin{figure}[h]
    \includegraphics[width=\linewidth]{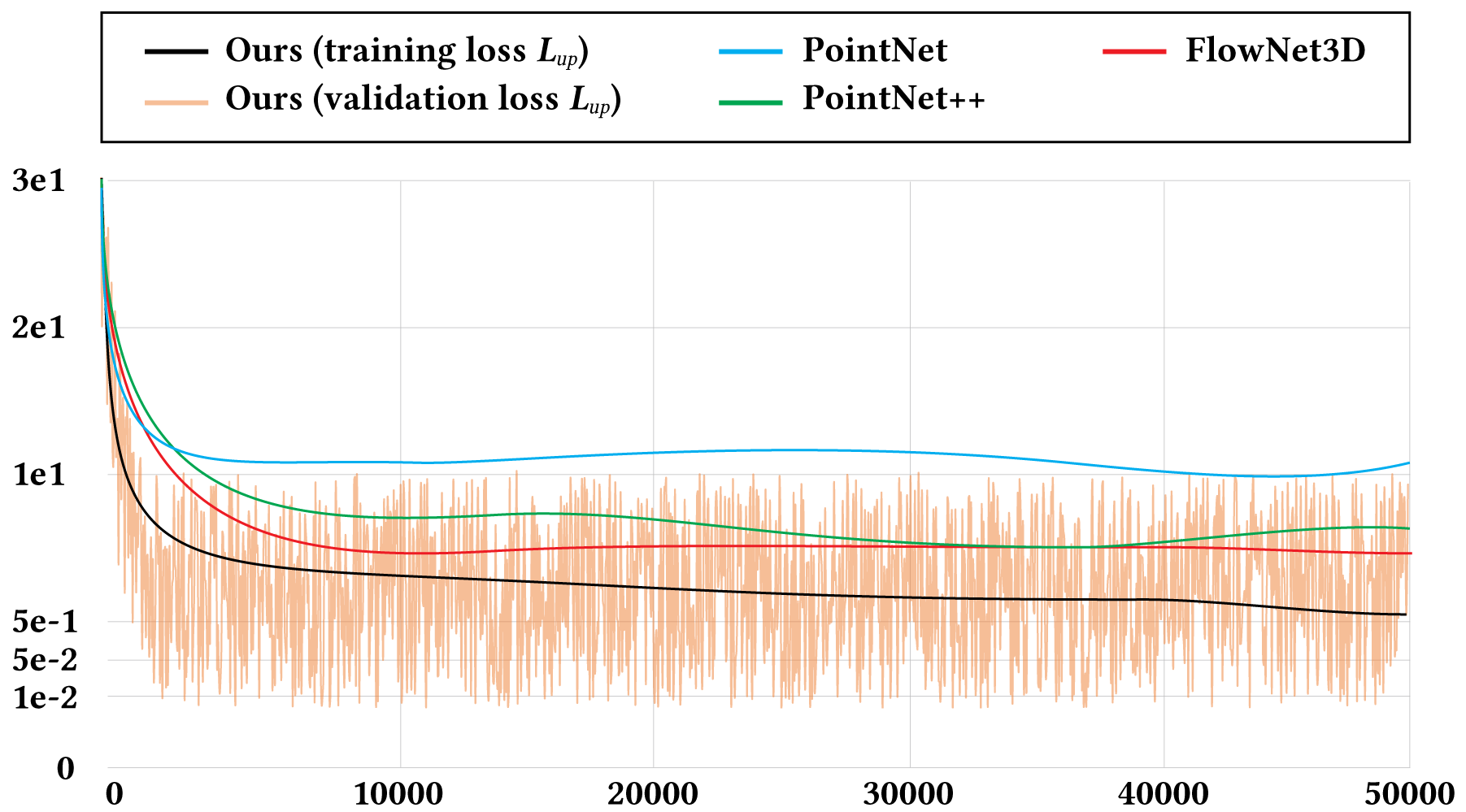}
    \caption{Comparing training losses on the \textsl{Container} dataset measured with our approach and a few point-based learning baselines. In order to lighten the chart, the validation loss (orange line) is showed only for our approach (black line).}
    \label{fig:loss_comparison}
\end{figure}

As highlighted in Table~\ref{table:compare_star}, our approach performs much better for the up-resing task on particle-based simulations when compared with the selected baselines. In addition to providing better results, our network converges much faster as shown in Fig.~\ref{fig:loss_comparison}. The convergence times are expressed in hours per epoch. Moreover, we have noticed that with our approach, up to 40\% of the EPE is due to static regions. These regions can show significant differences in volume when varying the simulation discretization parameters (i.e., particle spacing and grid resolution). When we manually exclude these regions (as suggested in \S~\ref{sec:upres_coarse_inputs}), the EPE reduces significantly.

\begin{figure}[h]
    \includegraphics[width=\linewidth]{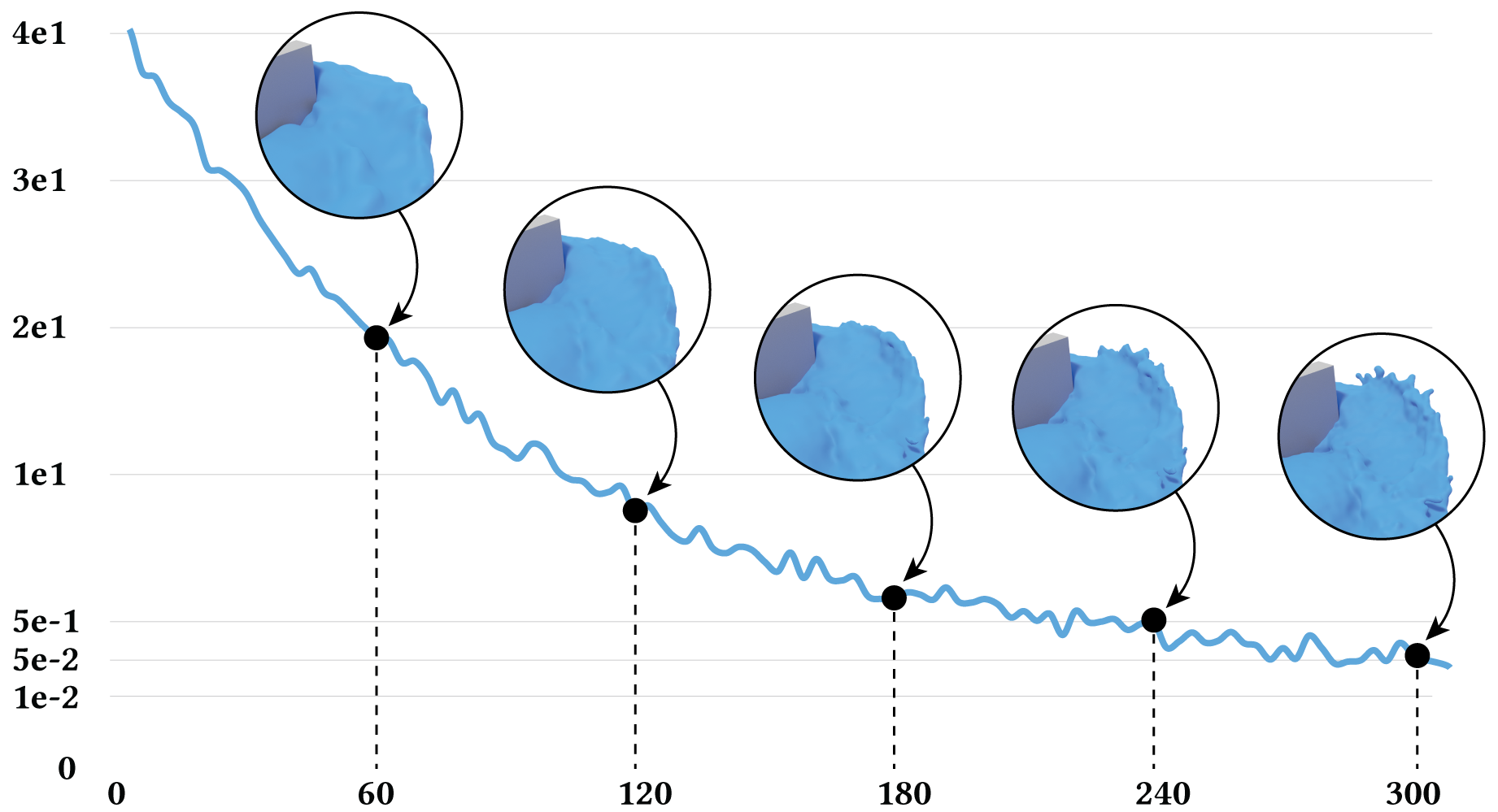}
    \caption{Convergence plot of our $L_{up}$ loss for the \textsl{Stirring} example from Fig.~\ref{fig:stiring_bowls}. We show the loss error ($Y$-axis) evolving throughout the training epochs ($X$-axis).}
    \label{fig:loss_error_epochs}
\end{figure}

Fig.~\ref{fig:loss_comparison} shows training losses evolving over iterations on several point-based learning methods. As shown, ours offers a more steady error reduction over iterations as opposed to the presented baselines. In other words, our approach requires fewer iterations to mimic the high-resolution details with good accuracy.
Fig.~\ref{fig:loss_comparison} also highlights the generalization capabilities on a way more complex example (Fig.~\ref{fig:acm_streams}). As expected, we also have observed that similar examples from the training set take less time to converge to a decent level of detail (as compared to the corresponding reference). We observed a clear up-resing displacement after around 200 epochs as shown in Fig.~\ref{fig:loss_error_epochs}. We noticed small differences in the precision of displacements with the range of 240 to 300 epochs (i.e., depending on the complexity of the input simulation), so for our purpose, training past that iteration threshold will unlikely bring significant value.

\subsection{Evaluation and Discussion}

\paragraph{\textbf{Up-Resing Coarse Input}}
\label{sec:upres_coarse_inputs}

In the presented results, we provide as input to our network the particle positions and a displacement field obtained between the current frame and the next one. We use these displacements to act as deformation preconditioners to our network (i.e., hint on spatial and temporal deformations). More importantly, using these preconditioners as deformations prevents us from computing an actual deformation field $\textbf{u}_{up}$ (i.e., using the UpFlOF algorithm) which would defy the objective of this approach by requiring a high-resolution input. The generated displacement by our network is then combined with the velocity of the coarse input to account for the coarse motion of the liquid since displacements are computed locally and with respect to their neighborhood.

\begin{figure}[h]
    \begin{subfigure}{0.3\linewidth}
        \includegraphics[width=\linewidth]{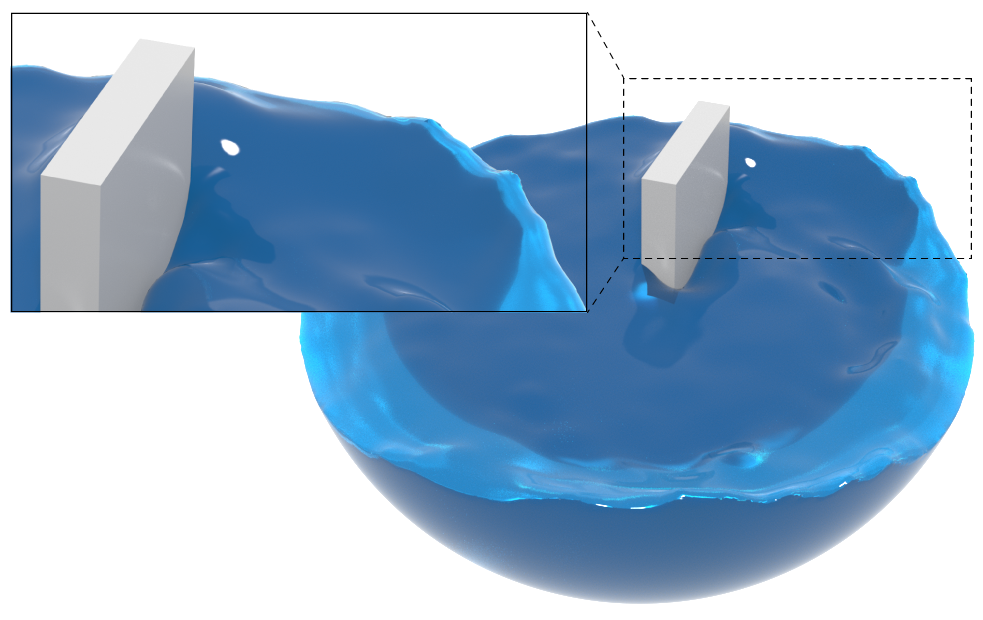}
        \caption{Coarse input}
        \label{fig:stiring_bowls_a}
    \end{subfigure}
    \begin{subfigure}{0.3\linewidth}
        \includegraphics[width=\linewidth]{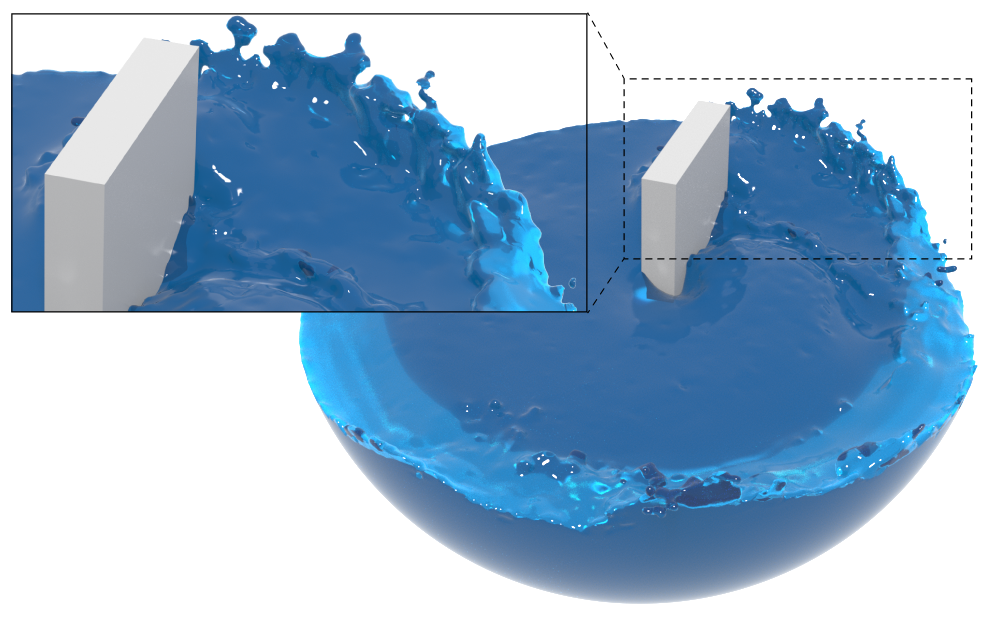}
        \caption{Ours}
        \label{fig:stiring_bowls_b}
    \end{subfigure}
    \begin{subfigure}{0.3\linewidth}
        \includegraphics[width=\linewidth]{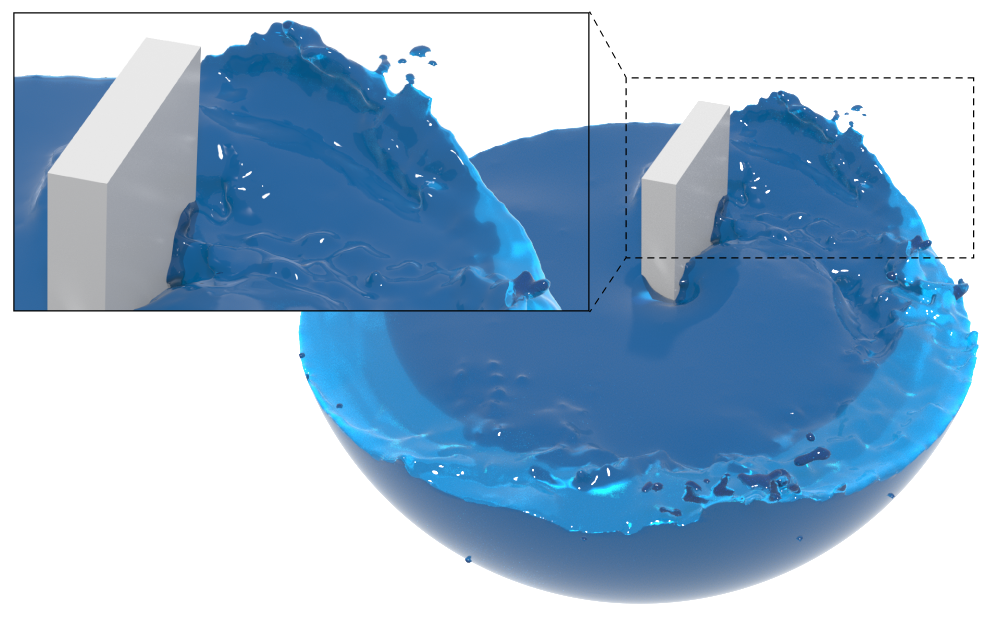}
        \caption{Reference}
        \label{fig:stiring_bowls_c}
    \end{subfigure}
    \caption{Comparison of our result (middle) generated with \textsl{FFNet} using a low-resolution input (left) with the corresponding high-resolution reference (right).}
    \label{fig:stiring_bowls}
\end{figure}

For convenience, we initially decided whether to process selected regions or the whole volume of liquid. The reason is that generating up-resing details on static volumes of liquid introduces noise due to the regression operations on downsampled neighborhoods.
\begin{wrapfigure}[8]{l}[0pt]{0.12\textwidth}
\vspace*{-0.5cm}
\includegraphics[width=0.15\textwidth]{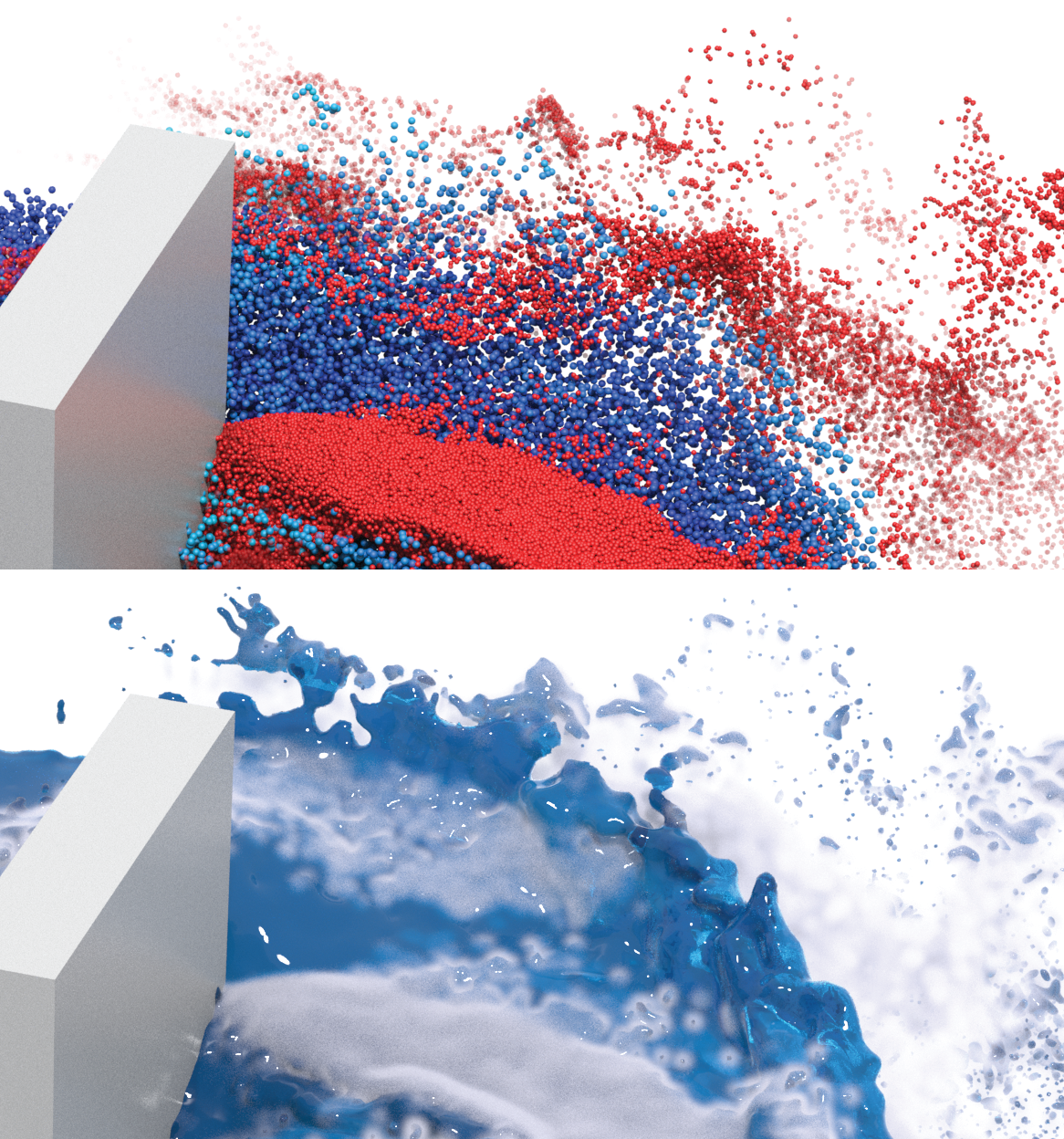} 
\label{fig:whitewater_parts}
\end{wrapfigure}
Therefore, in some of our examples, we initially flagged regions (and contained particles) that might be of interest to provide compelling details often absent in coarse simulations. However, in some cases, it was simpler to predict displacements for the whole simulation domain and then smooth out selected regions of the generated surface.

Lastly, it was observed that droplets and highly diffuse volumes of liquid are not faithfully generated by our displacement network. One solution could be to flag diffuse particles as proposed by Um et al.~\shortcite{um2018liquid}. For example, as exposed in the close-up insets of Fig.~\ref{fig:stiring_bowls}, our approach generates convincing splashing details while mostly remaining attached to the main volume of liquid. On the other hand, our network successfully reproduced eddies surrounding the moving obstacle stirring the liquid. However, as compared with the high-resolution reference (Fig.~\ref{fig:stiring_bowls_c}), we were unable to recreate the small droplets detached from the splashing portions of the liquid. Nevertheless, these missing droplets can easily be procedurally added (as shown with the red particles in the inlet figure) on top of our displaced particles (e.g., using existing tools such as Maya\textregistered~to create whitewater details).

Fig.~\ref{fig:teaser} is also a good example where adding spray and foam particles would improve the end results generated by our approach as it clearly lacks droplets compared to the high resolution reference. On a related note, we have noticed that sparse neighborhoods require more iterations during training to converge and to reconstruct these small-scale details. In other words, these diffuse regions contribute less throughout iterations and would require more epochs during training with our network.

\begin{figure*}[t]
    \begin{subfigure}{0.24\linewidth}
        \includegraphics[width=\linewidth]{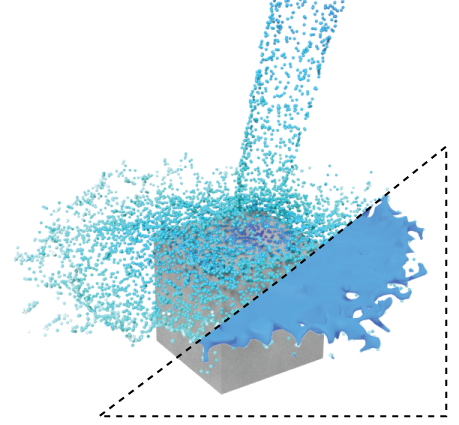}
        \caption{$d=R$}
    \end{subfigure}
    \begin{subfigure}{0.24\linewidth}
        \includegraphics[width=\linewidth]{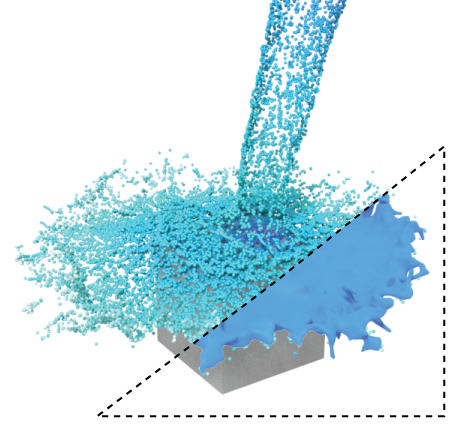}
        \caption{$d=0.75R$}
    \end{subfigure}
    \begin{subfigure}{0.24\linewidth}
        \includegraphics[width=\linewidth]{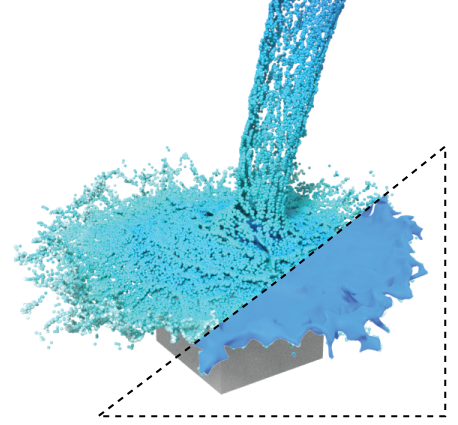}
        \caption{$d=0.5R$}
    \end{subfigure}
    \begin{subfigure}{0.24\linewidth}
        \includegraphics[width=\linewidth]{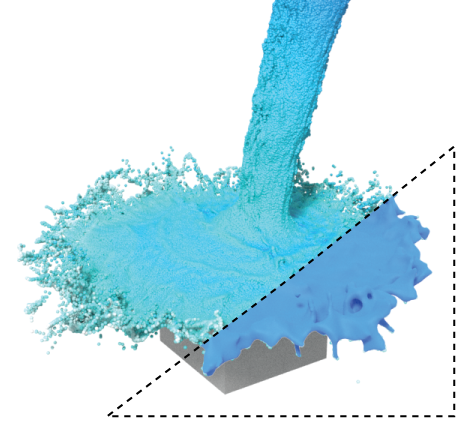}
        \caption{$d=0.25R$}
    \end{subfigure}
    \caption{We show the influence of upsampling the particle band prior to the inference step. As shown in this figure, we qualitatively evaluated the results by varying the resampling radius of the narrow band.}
    \label{fig:resampling_inference}
\end{figure*}

\paragraph{\textbf{Generalization on More Complex Inputs}}

\begin{figure}[h]
    \begin{subfigure}{0.4\linewidth}
        \includegraphics[width=\linewidth]{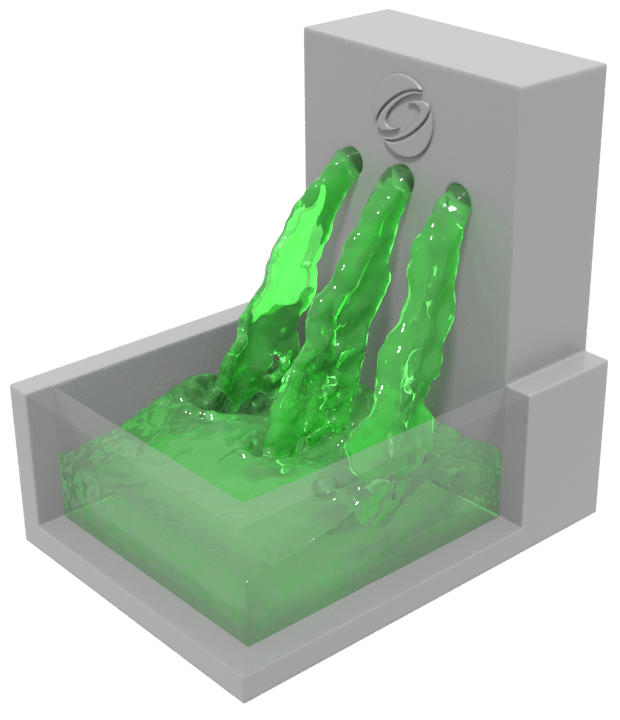}
        \caption{Coarse input}
        \label{fig:acm_streams_a}
    \end{subfigure}
    \begin{subfigure}{0.4\linewidth}
        \includegraphics[width=\linewidth]{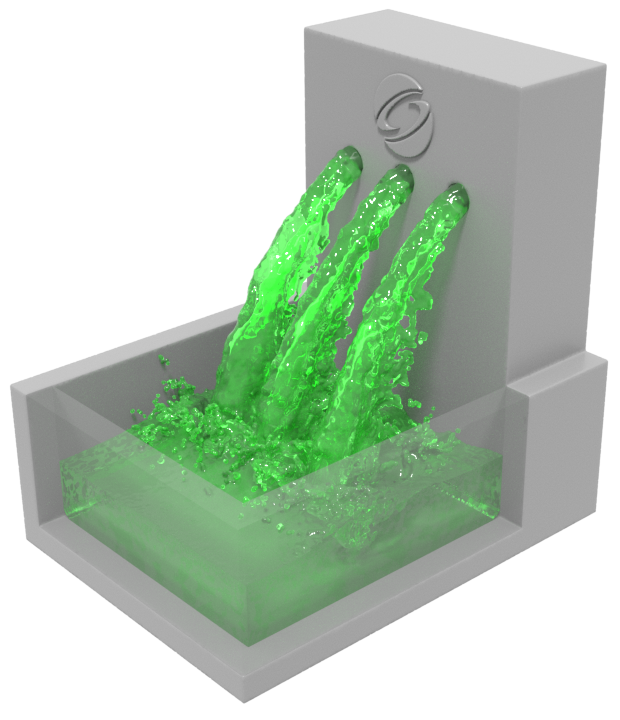}
        \caption{Reference}
        \label{fig:acm_streams_b}
    \end{subfigure}
    \begin{subfigure}{0.6\linewidth}
        \includegraphics[width=\linewidth]{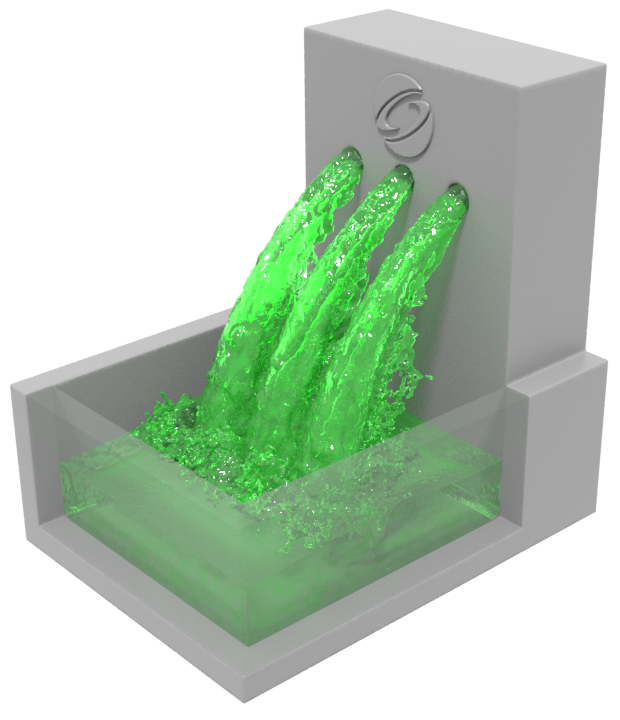}
        \caption{Ours}
        \label{fig:acm_streams_c}
    \end{subfigure}
    \caption{Example highlighting the generalization capacities of our network in an unknown simulation setup.}
    \label{fig:acm_streams}
\end{figure}

We have also tested the generalization capabilities of the proposed approach on more complex examples to evaluate how well it performs on unknown simulation setups. As exposed in \S~\ref{sec:datasets}, our dataset is composed of solely single sources of disturbance either on static liquid or static obstacles. In Fig.~\ref{fig:acm_streams}, we compare our results with the ground truth on a multiple streams example pouring into a single container. There are a few interesting features to observe in that figure. First, the small-scale details noticeable in the reference were successfully reproduced around the streams and at the impact points in the static liquid container. As previously mentioned, although our approach does not capture the diffuse particles very well (e.g., droplets), we can still observe a few detached chunks of liquid at the stream impacts.

\begin{figure}[H]
    \begin{subfigure}{0.3\linewidth}
        \includegraphics[width=\linewidth]{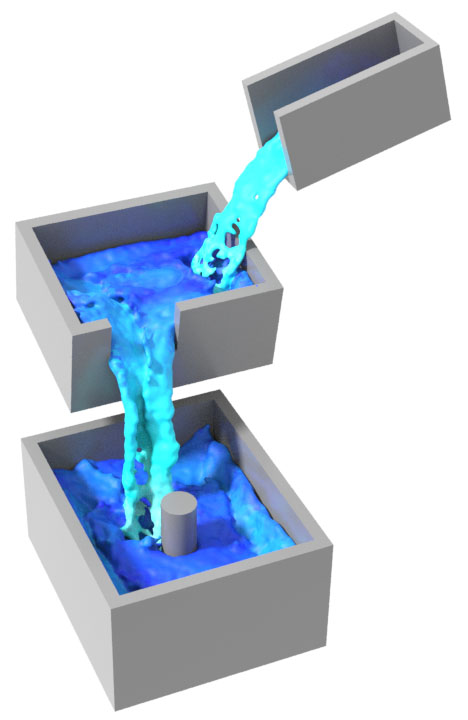}
        \caption{Coarse input}
        \label{fig:multi_stage_streams_a}
    \end{subfigure}
    \begin{subfigure}{0.3\linewidth}
        \includegraphics[width=\linewidth]{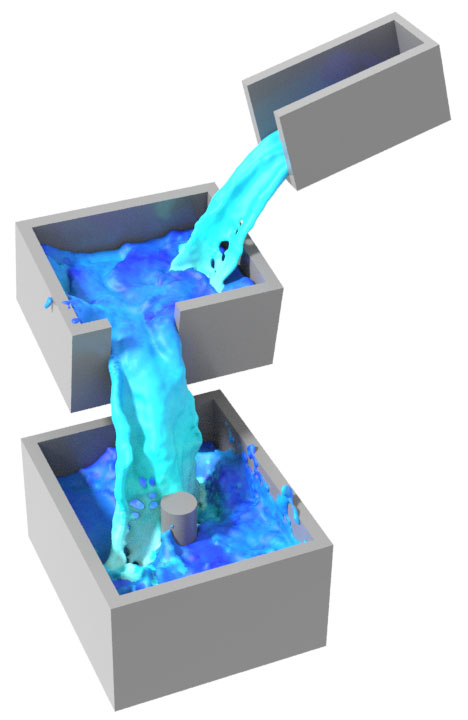}
        \caption{Ours}
    	\label{fig:multi_stage_streams_b}
    \end{subfigure}
    \begin{subfigure}{0.3\linewidth}
        \includegraphics[width=\linewidth]{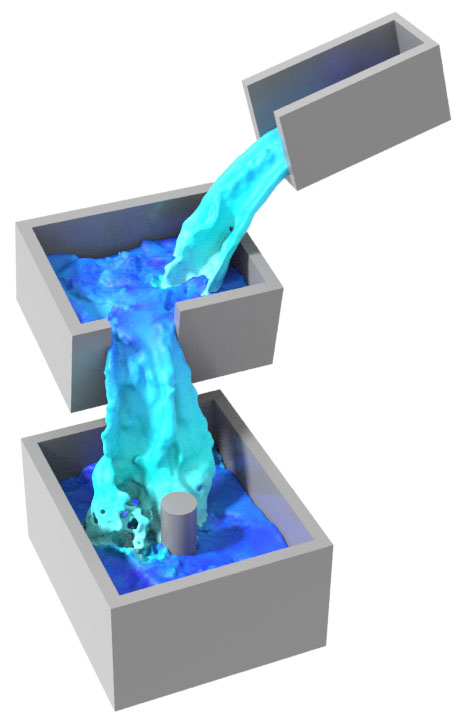}
        \caption{Reference}
    	\label{fig:multi_stage_streams_c}
    \end{subfigure}
    \caption{Comparison of our result (middle) generated with \textsl{FFNet} using a low-resolution input (left) with the corresponding high-resolution reference (right).}
    \label{fig:multi_stage_streams}
\end{figure}

Another interesting aspect observed in our results is the ability to reproduce the energy level of the simulations at higher resolutions. As noticeable in Fig.~\ref{fig:acm_streams}, the energy loss caused by a coarse discretization (Fig.~\ref{fig:acm_streams_a}) is partially restored in appearance when up-resed using our approach (Fig.~\ref{fig:acm_streams_c}). The arced streamlines in our result (Fig.~\ref{fig:acm_streams_c}) closely reproduce those of the high-resolution reference (Fig.~\ref{fig:acm_streams_b}). It is also apparent in Fig.~\ref{fig:multi_stage_streams} that the static container is significantly more agitated in our result than it is in the coarse input simulation. At each level of that cascading stream, our approach was able to infer dynamic and turbulent behavior within impacted regions. In the lower container (i.e., the one with the cylinder obstacle), we have noticed that our result (Fig.~\ref{fig:multi_stage_streams_b}) was showing a slightly different but plausible turbulence behavior (Fig.~\ref{fig:multi_stage_streams_c}).

\begin{figure}[h]
    \includegraphics[width=\linewidth]{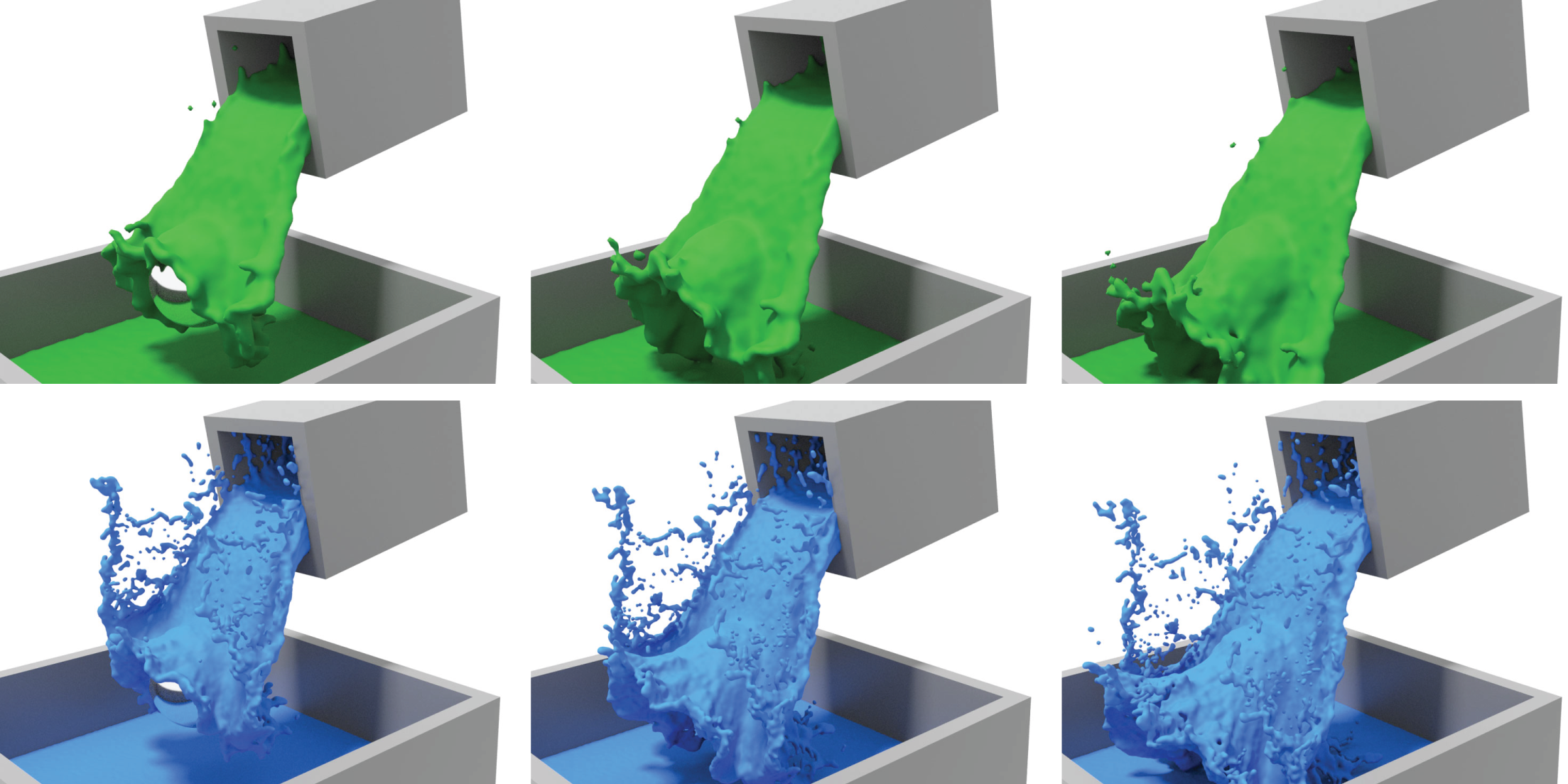}
    \caption{Comparing the evolution between our inferred liquid~(top) and the high-resolution reference~(bottom).}
    \label{fig:align_compare_ground_truth}
\end{figure}

\paragraph{\textbf{Improved Alignment}}
The alignment term (Eq.~\ref{eq:discretized_of_align}) is used to improve the precision of the deformation between resolutions. In order to fully reproduce high-resolution details on coarse liquids, the deformations applied with the inferred displacements of our network must be aligned in space and time to capture the small differences between the simulation resolutions. As an example, the impacts on the sphere shown in Fig.~\ref{fig:teaser} present noticeable differences between the coarse and the high-resolution liquids. Our goal with that alignment in that particular example would be to ensure that we are able to capture the detailed splashing impacts of the reference and infer this on the coarser liquid. With the exception of the detached droplets, we show in Fig.~\ref{fig:align_compare_ground_truth} that our approach (bottom row in green) does a fairly good job at reproducing similar fine details around the impact crown as compared to the reference (top row in blue).

\begin{figure}[h]
    \begin{subfigure}{0.48\linewidth}
        \includegraphics[width=\linewidth]{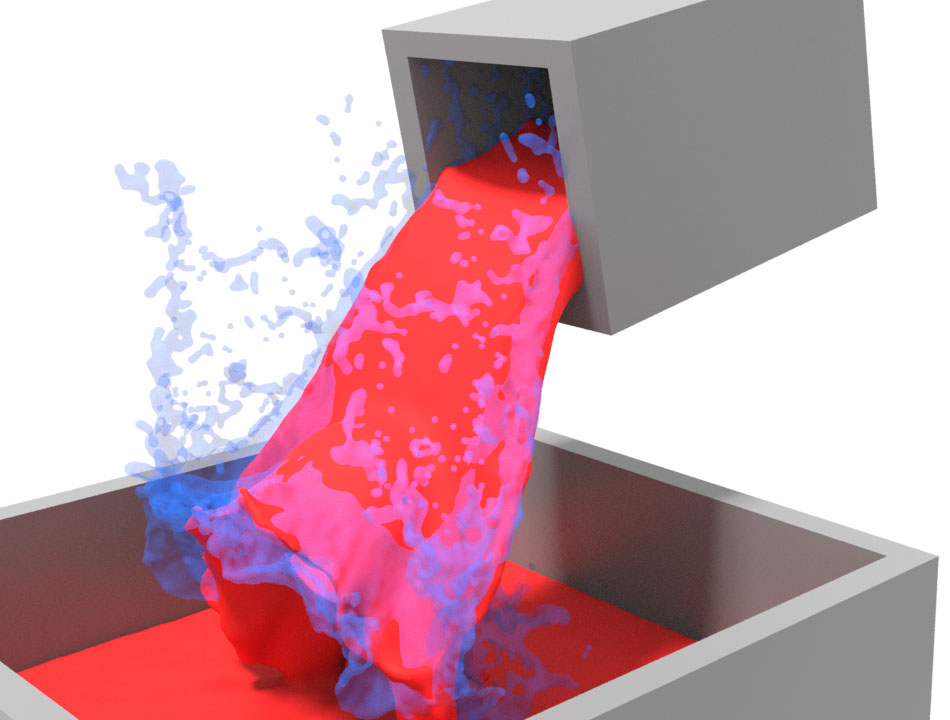}
        \caption{Without key-event alignment}
    \end{subfigure}
    \begin{subfigure}{0.48\linewidth}
        \includegraphics[width=\linewidth]{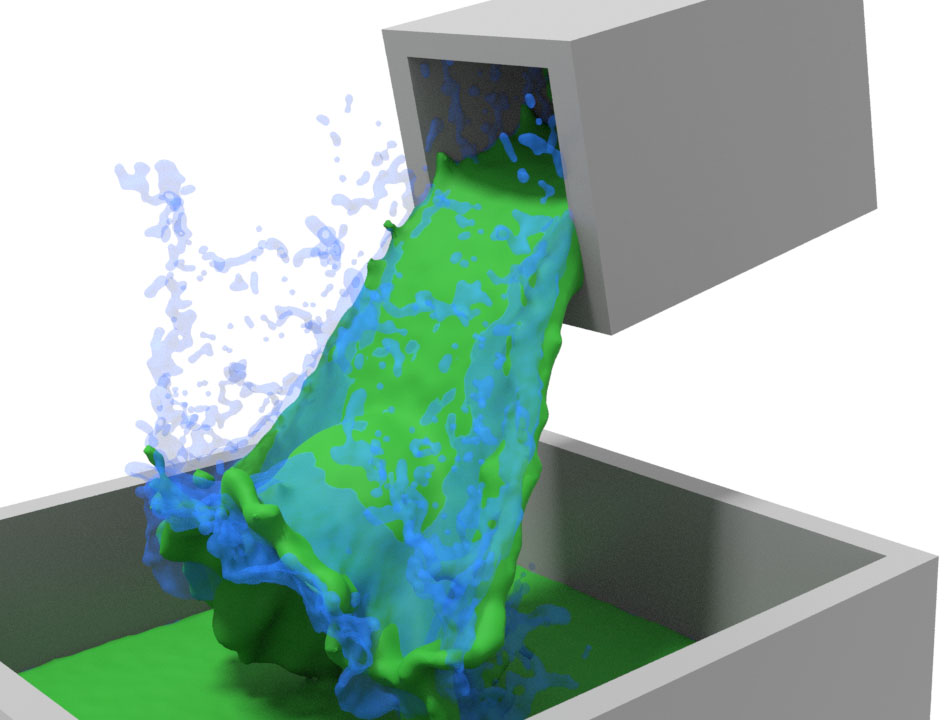}
        \caption{With $\textbf{D}$}
    \end{subfigure}
    \caption{Comparing before and after applying the proposed key-event alignment term on the teaser example (Fig.~\ref{fig:teaser}).}
    \label{fig:compare_with_alignment_constraint}
\end{figure}

The benefit of using key-event alignment is illustrated in Fig.~\ref{fig:compare_with_alignment_constraint}. In this figure, we compare the inter-resolution deformations applied with and without the alignment term $\textbf{D}$. We can observe that the green liquid fits better to the fine splashes of the reference (semi-transparent in blue) in comparison to the red liquid which presents the deformations before the correction on the alignment. The red liquid is missing most of the small-scale details of the main volume of liquid; this is due to a misalignment of the deformation field mapping the high-resolution to the low-resolution input.

\paragraph{\textbf{Upsampling for Inference}}
The upsampling process of an input liquid prior to inference is a crucial step to fully appreciate the details generated by our approach. By analogy, this step is comparable to the need to have enough material to model fine details on a physical object. In our context, the material is discretized as particles. Applying our approach directly to a coarse liquid (i.e., without oversampling) would result in an extremely diffuse liquid. The reason is as previously stated; modeling so much detail through our inferred displacements on a coarse liquid would only spread the particles sparingly. In addition, an insufficient particle density may cause unwanted artefacts on the surface of the up-resed liquid. Furthermore, it is important to note that the input resolution at inference determines the displacement resolution of our network’s output. In other words, the input resolution constrains the resolution of the inferred displacement. Nevertheless, the neighborhoods used by our convolution operators enable us to decouple our network from the input resolution while reproducing appealing details. Results shown in Figs.~\ref{fig:multi_stage_streams} and~\ref{fig:acm_streams} are concrete examples of cases in which the upsampling of the input simulation made it possible to faithfully reproduce several apparent characteristics in the associated reference.

In Fig.~\ref{fig:resampling_inference}, we show the influence of different distances $d$ over the precision of the inferred displacements. As shown in that figure, the smaller the sampling radius, the higher the level of detail of the applied displacements. We have used $d=0.5R$ in most of the examples presented to limit the size of the input liquid to our network and because using a larger distance did not have a noticeable difference in the final result. Lastly, similarly to the work of Liu et al.~\shortcite{liu2019flownet3d}, we also perform multiple passes of inference (usually 3-6 passes) using different upsampling distances of the surface band to reduce the inference noise and to prevent particles from emerging following the application of the displacements.



\section{Conclusion and Future Work}
We have presented an approach leveraging deep learning to increase the apparent resolution of a coarse particle-based liquid. In addition, we have proposed a framework using a state-of-the-art interpolation method to generate and augment a dataset of particle-based simulations for machine learning purpose. Our approach can infer plausible and complex details on the surface of low-resolution liquids, as illustrated in the examples in this paper and animated sequences in the accompanying video.

Looking toward the future, we believe that using a stacked network could improve our results with highly diffuse liquids and thin sheets. Our work could be combined with a spray particle classification network such as proposed by Um et al.~\shortcite{um2018liquid} to update an adaptive neighborhood within the convolution layers. Finally, we think that it would also be interesting to investigate volumetric learning methods targeting up-resing applications like this one but for smoke simulations.

\begin{acks}
This work was supported and funded by Mitacs Accelerate, Mitacs Globalink, the Université de Montréal, and the École de Technologie Supérieure (ÉTS). We would also like to give a special thanks to Nils Thuerey for his insightful advice, comments, and discussions surrounding this project during the research internship at the Technical University of Munich (TUM) at the very beginning of this project. Finally, We would like to thank Autodesk, Inc. for providing resources to carry out this project in the end.
\end{acks}

\bibliographystyle{ACM-Reference-Format}
\bibliography{arXiv_paper}

\end{document}